\documentclass[10pt,twocolumn,twoside]{IEEEtran}

\pdfoutput=1

\usepackage[cmex10]{amsmath}
\usepackage{cite}

\usepackage{latexsym}
\usepackage{graphics}
\usepackage{amssymb}
\usepackage{amsthm}

\usepackage{cite}
\usepackage{array}
\usepackage{mdwmath}
\usepackage{mdwtab}
\usepackage[tight,footnotesize]{subfigure}
\usepackage{graphicx}

\usepackage{stfloats}
\fnbelowfloat
\usepackage{url}

\ifCLASSOPTIONcompsoc
\usepackage[tight,normalsize,sf,SF]{subfigure}
\else
\usepackage[tight,footnotesize]{subfigure}
\fi

\def\vectorize{\mathrm{vec}}

\def\kron{\otimes}
\def\tr{\mathrm{tr}}

\def\spa{\mathrm{span}}

\newcommand{\argmax}[1]{{\underset{{#1}}{\mathrm{arg\,max}}}}
\newcommand{\argmin}[1]{{\underset{{#1}}{\mathrm{arg\,min}}}}
\newcommand{\fracSum}[1]{{\underset{{#1}}{\sum}}}

\newcommand{\condProd}[3]{\overset{#3}{\underset{\underset{#2}{#1}}{\prod}}}
\newcommand{\condProdtwo}[2]{\overset{#2}{\underset{#1}{\prod}}}
\newcommand{\fracSumtwo}[2]{\overset{#2}{\underset{#1}{\sum}}}
\newcommand{\vect}[1]{\mathbf{#1}}

\newcommand{\maximize}[1]{{\underset{{#1}}{\mathrm{maximize}}}}

\theoremstyle{remark}
\newtheorem{remark}{Remark}

\newtheorem{theorem}{Theorem}
\newtheorem{corollary}{Corollary}
\newtheorem{lemma}{Lemma}
\newtheorem{definition}{Definition}

\begin{document}

\title{Receive Combining vs.~Multi-Stream Multiplexing in Downlink Systems with
Multi-Antenna Users}

\author{Emil Bj\"ornson,~\IEEEmembership{Member,~IEEE,}
Marios Kountouris,~\IEEEmembership{Member,~IEEE,}
       Mats Bengtsson,~\IEEEmembership{Senior Member,~IEEE,}
     and~Bj\"orn~Ottersten,~\IEEEmembership{Fellow,~IEEE}
\thanks{\copyright 2013 IEEE. Personal use of this material is permitted. Permission from IEEE must be obtained for all other uses, in any current or future media, including reprinting/republishing this material for advertising or promotional purposes, creating new collective works, for resale or redistribution to servers or lists, or reuse of any copyrighted component of this work in other works.}%
\thanks{Supplementary downloadable material is available at https://github.com/emilbjornson/one-or-multiple-streams, provided by the authors. The material includes Matlab code that reproduces all simulation results.}%
\thanks{The research leading to these results has received funding from the European Research
Council under the European Community's Seventh Framework Programme
(FP7/2007-2013)/ERC grant agreement number 228044. The work of E.~Bj\"ornson is funded by the International Postdoc Grant 2012-228 from The
Swedish Research Council. This work was presented in part at the IEEE
Swedish Communication Technologies Workshop (Swe-CTW), Stockholm,
Sweden, October 2011 [32].}%
\thanks{E.~Bj\"ornson, M.~Bengtsson,
and B.~Ottersten are with the Signal Processing Laboratory, ACCESS
Linnaeus Center, KTH Royal Institute of Technology, SE-100 44
Stockholm, Sweden (e-mail: emil.bjornson@ee.kth.se;
mats.bengtsson@ee.kth.se; bjorn.ottersten@ee.kth.se). B.~Ottersten is also with Interdisciplinary Centre for Security, Reliability and
Trust (SnT), University of Luxembourg, L-1359 Luxembourg-Kirchberg, Luxembourg (email: bjorn.ottersten@uni.lu). M.~Kountouris and E.~Bj\"ornson are with SUPELEC (Ecole Sup$\acute{\textrm{e}}$rieure d'Electricit$\acute{\textrm{e}}$), Gif-sur-Yvette, France (e-mail: marios.kountouris@supelec.fr;  emil.bjornson@supelec.fr).}%
}

\markboth{IEEE TRANSACTIONS ON SIGNAL PROCESSING, VOL.~61, NO.~13, JULY 1, 2013}%
{Bj\"ornson \MakeLowercase{\textit{et al.}}: RECEIVE COMBINING VS.~MULTI-STREAM MULTIPLEXING IN DOWNLINK SYSTEMS WITH MULTI-ANTENNA USERS}

\maketitle

\begin{abstract}
In downlink multi-antenna systems with many users, the multiplexing gain is strictly limited by the number of transmit antennas $N$ and the use of these antennas. Assuming that the total number of receive antennas at the multi-antenna users is much larger than $N$, the maximal multiplexing gain can be achieved with many different transmission/reception strategies. For example, the excess number of receive antennas can be utilized to schedule users with effective channels that are near-orthogonal, for multi-stream multiplexing to users with well-conditioned channels, and/or to enable interference-aware receive combining. In this paper, we try to answer the question if the $N$ data streams should be divided among few users (many streams per user) or many users (few streams per user, enabling receive combining). Analytic results are derived to show how user selection, spatial correlation, heterogeneous user conditions, and imperfect channel acquisition (quantization or estimation errors) affect the performance when sending the maximal number of streams or one stream per scheduled user---the two extremes in data stream allocation.

While contradicting observations on this topic have been reported in prior works, we show that selecting many users and allocating one stream per user (i.e., exploiting receive combining) is the best candidate under realistic conditions. This is explained by the provably stronger resilience towards spatial correlation and the larger benefit from multi-user diversity. This fundamental result has positive implications for the design of downlink systems as it reduces the hardware requirements at the user devices and simplifies the throughput optimization.
\end{abstract}


\IEEEpeerreviewmaketitle

\section{Introduction}

The performance of downlink wireless communication systems can be improved by multi-antenna techniques, which enable efficient utilization of spatial dimensions. Depending on the available channel state information (CSI), these dimensions can be used for enhanced reliability and/or spatial multiplexing of multiple data streams with controlled interference \cite{Gesbert2007a}.
The downlink single-cell sum capacity (with perfect CSI) behaves as
\begin{equation} \label{eq_sumcapacity}
\min(N,MK) \log_2(P) + \mathcal{O}(1)
\end{equation}
where $N$ is the number of base station antennas, $K$ is the number of users, each user has $M \geq 1$ antennas, and $P$ is the signal-to-noise ratio (SNR) defined as the
total transmit power divided by the noise power.
The number of users is assumed to be large such that $K\geq N$, thus we have $MK \geq N$ and the maximal \emph{multiplexing gain} becomes $\min(N,MK)=N$.
The multiplexing gain will have a major impact on the throughput of future cellular networks, where high SNRs can be achieved in an energy-efficient way by large-scale antenna arrays \cite{Rusek2013a} and/or increased cell density \cite{Hoydis2011c}.

The sum capacity in \eqref{eq_sumcapacity} is theoretically achieved by dirty-paper coding \cite{Weingarten2006a}, but this non-linear scheme has impractical complexity and is very sensitive
to CSI imperfections. Fortunately, the maximal multiplexing gain of $N$ can be achieved by linear \emph{spatial division multiple access} (SDMA) strategies \cite{Richard1996a}, such as \emph{block-diagonalization (BD)} \cite{Spencer2004a,Ravindran2008a} and \emph{zero-forcing with combining (ZFC)} \cite{Jindal2008a,Trivellato2008a}. Such SDMA strategies transmit $N$ simultaneous data streams, but can divide them among the users in different ways; the system can select between $\lceil \frac{N}{M} \rceil$ and $N$ users to be active and allocate from $1$ to $M$ streams to each of them. This raises a fundamental design question: \emph{how should the receive antennas at each user be used to maximize the system throughput?}

Inter-user interference degrades user performance, while the mutual interference between users' own streams can be handled by receive processing.
It thus seems beneficial to only have a few active users and multiplex many streams to each of them. However, one should keep in mind that every additional stream allocated to a user
experiences a weaker channel gain than the previous streams.
If fewer than $M$ streams are allocated to a user, this user has degrees of freedom for interference-aware receive combining to achieve a strong effective channel and better spatial co-user compatibility. In other words, it is not clear whether receive antennas should be utilized for \emph{multi-stream multiplexing} or \emph{receive combining}, or perhaps something intermediate.
The answer has a profound impact on wireless system design, including the CSI acquisition protocols, scheduling algorithms, and receiver architecture.

\subsection{Related Work}

The sum-rate maximization problem is nonconvex and combinatorial \cite{Luo2008a}, thus only suboptimal strategies are feasible in practice. Such low-complexity algorithms have been proposed in \cite{Tolli2005a,Boccardi2007a,Chen2008a,Guthy2010a}, among others, by successively allocating data streams to users in a greedy manner. Simulations have indicated that fewer than $N$ streams should be used when $P$ and $K$ are small, and that spatial correlation makes it beneficial to divide the streams among many users.
Simulations in \cite{Boccardi2007a} indicates that the probability of allocating more than one stream per user is small when $K$ grows large, but \cite{Boccardi2007a} only considers users with homogeneous channel conditions and all the aforementioned papers assume perfect CSI.

The authors of \cite{Trivellato2008a} claim that transmitting at most one stream per user is desirable when there are many users in the system. They justify this statement by using asymptotic results from \cite{Bayesteh2008a} where $K\rightarrow \infty$. This argumentation ignores some important issues: 1) asymptotic optimality can also be proven with multiple streams per user;\footnote{The uplink analysis in \cite{Rhee2004a} shows that a non-zero (but bounded) number of users can use multiple streams, and the well-established uplink-downlink duality makes this result applicable also in our downlink scenario.} 2) the performance at practical values on $K$ is unknown; and 3) the analysis implies an unbounded asymptotic multi-user diversity gain, which is a modeling artifact of fading channels \cite{Dohler2011a}.

The authors of \cite{Jindal2008a,Ravindran2008a} arrive at a different conclusion when they compare BD (which selects $\frac{N}{M}$ users and sends $M$ streams/user) and ZFC (which selects $N$ users and sends one stream/user) under quantized CSI. Their simulations reveal a distinct advantage of BD (i.e., multi-stream multiplexing), but are limited to uncorrelated channels and neither include user selection nor interference rejection. We show that their results are misleading, because  single-user transmission greatly outperforms both BD and ZFC in the scenario that they simulate.

Despite the similar terminology, our problem is fundamentally different from the classic works on the diversity-spatial multiplexing tradeoff (DMT) in \cite{Zheng2003a,Lozano2010a}. The DMT brings insight on how many streams should be transmitted in the high-SNR regime, while we consider how a fixed number of streams should be divided among the users.

\subsection{Main Contributions}

\begin{figure}
\begin{center}
\subfigure[1 stream per 2-antenna user: ZFC enables receive combining.]
{\label{figure_combining}\includegraphics[width=86mm]{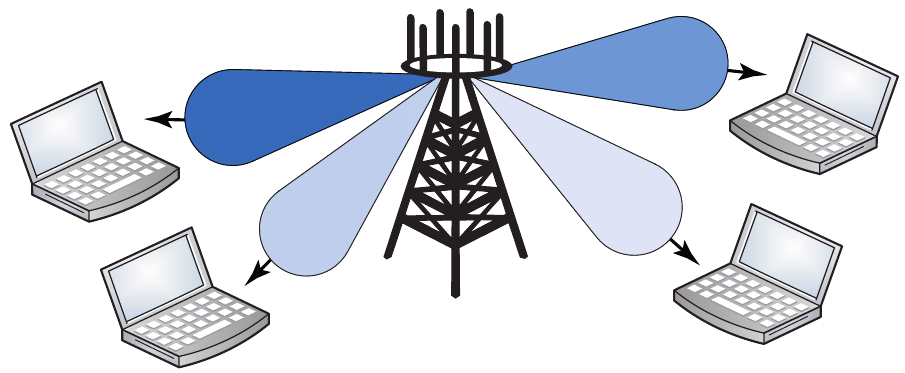}}\hfill
\subfigure[2 streams per 2-antenna user: BD exploits multi-stream multiplexing.]
{\label{figure_multiplexing}\includegraphics[width=86mm]{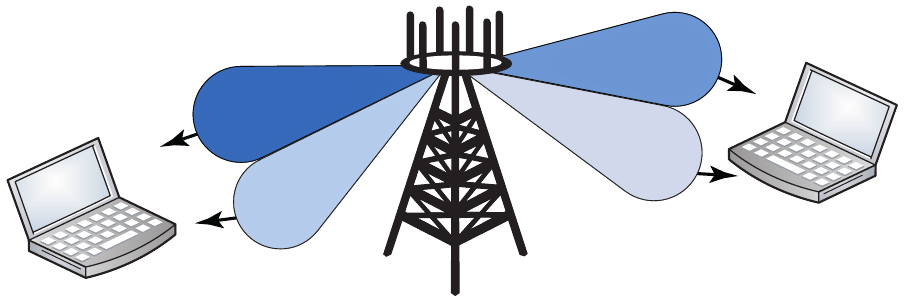}} \vskip-2mm
\caption{Two ways of dividing four data streams among multi-antenna users, which also represents two ways of utilizing the receive antennas to reduce interference.
(a) Receive one stream per user and linearly combine the antenna to achieve an effective channel that rejects interference. (b) Receive multiple streams and handle their mutual interference through receive processing.}\label{figure_combining_vs_multiplexing}
\end{center} \vskip-5mm
\end{figure}

This paper provides a comprehensive answer to how multi-antenna users should utilize their antennas in downlink transmissions, or similarly how many data streams that should be allocated per active user under different system conditions; see Fig.~\ref{figure_combining_vs_multiplexing}. The main contributions are:

\begin{itemize}
\item New analytic results for analyzing the problem under spatial correlation, user selection, heterogeneous user channel conditions, and realistic CSI acquisition.
These enable asymptotic comparison of the two extremes: allocating $M$ streams per active user (called BD) and one stream per active user (called ZFC). We show that ZFC is more resilient to spatial correlation and well adapted to find near-orthogonal users, while BD is better at utilizing heterogeneous user conditions. Imperfect CSI acquisition is shown to have a similar impact on both strategies.

\item Numerical examples show that allocating one stream per active user is essentially optimal under realistic system conditions, and we explain how other conclusions may arise. The main conclusion is that utilizing receive combining is preferable over multi-stream multiplexing.
\end{itemize}

\section{System Model}

We consider a downlink multi-user MIMO system where a single base station with $N$ antennas communicates with $K \geq N$ users.  Each user has $M$ antennas. For analytical convenience\footnote{The case $M \geq N$ is analytically different because 1) Single-user transmission achieves the full multiplexing gain; 2) CSI acquisition is simplified since $\vect{H}_k$ has full row rank, thus any effective channel $\vect{C}_k^H \vect{H}_k$ can be achieved by selecting the receive combining $\vect{C}_k$ properly. Since user devices are size-constrained, the case $M < N$ is also reasonable in practice.} we assume that $M < N$ and often also that $\frac{N}{M}$ is an integer, but the precoding strategies considered herein can be applied for any $M$. The narrowband, flat-fading channel to user $k$ is represented in the complex-baseband by $\vect{H}_k \in \mathbb{C}^{M \times N}$. The received signal at this user is
\begin{equation} \label{eq_system_model}
\vect{y}_k = \vect{H}_k \vect{x} + \vect{n}_k
\end{equation}
where $\vect{x} \in \mathbb{C}^{N \times 1}$ is the joint transmitted signal for all users and $\vect{n}_k \sim \mathcal{CN}(\vect{0},\vect{I}_{M})$ is the (normalized) circularly-symmetric complex Gaussian noise vector. For analytic convenience, and motivated by measurements \cite{Chizhik2003a,Yu2004b}, we employ the Kronecker model with $\vect{H}_k = \vect{R}_{R,k}^{1/2} \widetilde{\vect{H}}_k \vect{R}_{T,k}^{1/2}$, where $\vect{R}_{T,k}$ and $\vect{R}_{R,k}$ are the positive-definite spatial correlation matrices at the transmitter and receiver side, respectively, and $\widetilde{\vect{H}}_k$ has independent $\mathcal{CN}(0,1)$-entries. We assume $\vect{R}_{T,k} = \vect{I}_N$ (i.e., large antenna separation at the base station) throughout the analysis, because transmit correlation
both creates complicated mathematical structures and requires limiting assumptions on the user distribution geometry and fading environment.
Observe that $\vect{R}_{R,k}$ generally is different for each users, describing different spatial properties.

\subsection{Cyclic System Operation}

\begin{figure}[t!]
\begin{center}
\subfigure[]
{\label{figure_fdd_operation}\includegraphics[width=86mm]{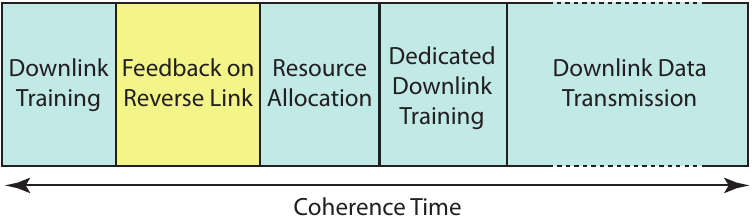}}\hfill
\subfigure[]
{\label{figure_tdd_operation}\includegraphics[width=86mm]{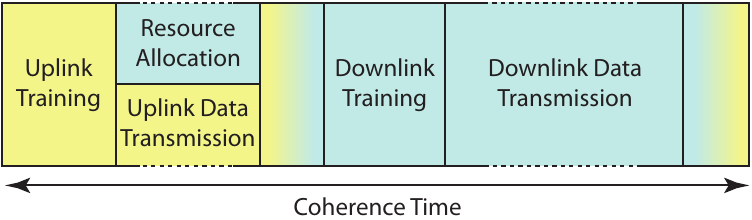}} \vskip-2mm
\caption{Basic block-fading system operation of (a) FDD systems; and (b) TDD systems. The system operation is repeated in a cyclic manner.}\label{figure_fdd_tdd}
\end{center} \vskip-5mm
\end{figure}

We assume block fading where $\vect{H}_k$ is static for a set of channel uses, called the \emph{coherence time}, and then updated independently. We consider both frequency division duplex (FDD) and time division duplex (TDD); baselines of the respective cyclic system operations are illustrated in Fig.~\ref{figure_fdd_tdd}.

In FDD systems, the users acquire CSI through training signaling \cite{Bjornson2010a} and some users feed back quantized CSI. The base station then performs resource allocation (i.e., data stream allocation and precoding) and informs the scheduled users of their precoding through a second training stage. Data transmission follows until the end of the coherence time, when the cycle in Fig.~\ref{figure_fdd_operation} restarts.

In TDD systems, the system toggles between uplink and downlink transmission on the same channel, thus enabling training signaling in both directions. We assume perfect channel reciprocity\footnote{The physical channel is always reciprocal, but different transceiver hardware is typically used in the downlink and the uplink. Thus, careful calibration is necessary to utilize the reciprocity in practice.} and that the coherence time makes CSI obtained in one block of Fig.~\ref{figure_tdd_operation} correct until the same block occurs in the next cycle. The base station does resource allocation for both uplink and downlink, and it informs the users through training signaling.

We assume that all training signals sent in the downlink direction provide the users with perfect CSI, while CSI feedback (in FDD) and uplink training (in TDD) might lead to imperfect CSI at the base station. This assumption enables coherent reception, thus making the conventional achievable sum rate expression a reasonable performance measure.\footnote{Many of the results herein can be extended to include imperfect CSI at the users in the resource allocation, followed by a second training stage that provides scheduled users with sufficiently accurate CSI of the precoded channels to enable coherent reception. See \cite{Caire2010a} for an example in FDD systems. The loss of having imperfect CSI also in the second training stage can be characterized as in \cite{Yoo2006b}.}

\subsection{Linear Precoding: General Problem Formulation}

\label{subsection_general_problem_formulation}

We consider linear precoding and the transmitted signal is
\begin{equation}
\vect{x} = \sum_{k=1}^{K} \vect{W}_k \vect{d}_k
\end{equation}
where $\vect{W}_k \in \mathbb{C}^{N \times d_k}$ is the precoding matrix, $\vect{d}_k \sim \mathcal{CN}(\vect{0},\vect{I}_{d_k})$ is the data signal, and $d_k$ is the number of multiplexed data streams to user $k$. Each user applies a semi-unitary receive combining matrix $\vect{C}_k \in \mathbb{C}^{M \times d_k}$ (i.e., $\vect{C}_k^H \vect{C}_k = \vect{I}_{d_k}$) and treats inter-user interference as Gaussian noise. The achievable information rate is
\begin{equation} \label{eq_data_rate_general}
g_k(\{\vect{W}_{\ell}\},\vect{C}_{k}) =  \log_2 \! \frac{\det \!\Big( \vect{I}_{d_k} \!+\! \fracSumtwo{\ell=1}{K} \vect{C}_k^H \vect{H}_k  \vect{W}_{\ell} \vect{W}_{\ell}^H \vect{H}_k^H \vect{C}_k \Big) \!}{ \det \!\Big( \vect{I}_{d_k} \!+\! \fracSum{\ell \neq k}  \vect{C}_k^H \vect{H}_k \vect{W}_{\ell} \vect{W}_{\ell}^H \vect{H}_k^H \vect{C}_k \Big) \! }
\end{equation}
where $\{\vect{W}_{\ell}\}$ denotes the set of precoding matrices and $\ell$ is an arbitrary user index \cite{Guthy2010a}. The transmission is limited by an average power/SNR constraint of $P$, thus
\begin{equation}
\mathbb{E}\{ \vect{x}^H \vect{x} \} = \sum_{k=1}^{K} \tr( \vect{W}_k \vect{W}_k^H ) \leq P.
\end{equation}

Ideally, we would like to select $\vect{W}_{k},\vect{C}_{k},d_k \, \forall k$ to maximize the sum rate; that is,
\begin{equation} \label{eq_sum_rate_early}
\begin{split}
\maximize{\{\vect{W}_{k},\vect{C}_{k},d_k \}} \,\, & \,\, \sum_{k=1}^{K} g_k(\{\vect{W}_{\ell}\},\vect{C}_{k}) \\
\mathrm{subject}\,\,\mathrm{to}\,\, & \, \sum_{k=1}^{K} \tr( \vect{W}_k \vect{W}_k^H ) \leq P, \\
& \, \vect{C}_k^H \vect{C}_k = \vect{I}_{d_k}, \quad \quad d_k\geq 0 \quad \forall k.
\end{split}
\end{equation}
Unfortunately, this resource allocation problem is NP-hard and therefore not practically solvable \cite{Luo2008a}.
There are algorithms that find local optima of \eqref{eq_sum_rate_early} (see \cite{Shi2011} and references therein), but these are iterative and thus cannot be implemented under the cyclic system operation in Fig.~\ref{figure_fdd_tdd}.

We limit the selection of $\{\vect{W}_{k},\vect{C}_{k},d_k \}$ to achieve a tractable problem formulation.

\begin{enumerate}
\item \textbf{Precoding:} Zero or minimal inter-user interference should be caused, which is possible when $\sum_{k=1}^{K} d_k \leq N$. This makes \eqref{eq_sum_rate_early} partially feasible, because it becomes a convex problem for any fixed $\vect{C}_{k},d_k$. This is a non-limiting assumption at high SNR \cite{Yoo2006a}, which is the regime where systems with high spectral efficiencies need to operate (e.g., using high power, small cells, or large antenna arrays \cite{Hoydis2011c,Rusek2013a}).

\item \textbf{Receive combining:} The matrix $\vect{C}_k$ is fixed at some value $\widetilde{\vect{C}}_k$ beforehand. This makes sense from a CSI acquisition perspective as only the effective channel $\widetilde{\vect{C}}_k^H \vect{H}_k$ needs to be obtained through feedback (in FDD) or training signaling (in TDD). The value $\widetilde{\vect{C}}_k$ might be the $d_k$ strongest (left) singular vectors of $\vect{H}_k$, known as maximum ratio combining (MRC), but can also be selected to improve the CSI feedback accuracy \cite{Jindal2008a,Trivellato2008a}.

\item \textbf{Stream allocation:} Users are scheduled sequentially using some predefined scheduling policy. This avoids making an exhaustive search over all data stream allocations, which is practically infeasible when $N$ and $K$ grow large. Greedy scheduling algorithms can perform remarkably close to optimum \cite{Tolli2005a,Boccardi2007a,Chen2008a,Guthy2010a}, while random selection ensures user fairness.
\end{enumerate}

We now have a simplified resource allocation problem,
\begin{equation} \label{eq_practical_resource_allocation}
\begin{split}
\maximize{\{\vect{W}_{k}\}} \,\, & \, \sum_{k \in \mathcal{S}} \log_2 \det ( \vect{I}_{d_k} +\widetilde{\vect{C}}_k^H \vect{H}_k \vect{W}_{k} \vect{W}_{k}^H \vect{H}_k^H \widetilde{\vect{C}}_k ) \\
\mathrm{subject}\,\,\mathrm{to}\,\, & \, \sum_{k \in \mathcal{S}} \tr( \vect{W}_k \vect{W}_k^H ) \leq P, \\
& \, \widetilde{\vect{C}}_k^H \vect{H}_k \vect{W}_{\ell} = \vect{0}_{d_k \times d_{\ell}} \quad \forall k \in \mathcal{S}, \, \forall \ell \in \mathcal{S} \!\setminus \! \{ k \},
\end{split}
\end{equation}
where $\mathcal{S}$ is the scheduling set given by the predefined scheduling rule and $d_k>0$ for $k \in \mathcal{S}$ is the corresponding data stream allocation.

\begin{remark}[Updating the Receive Combiner] \label{remark:receive-combiner}
When \eqref{eq_practical_resource_allocation} has been solved, the users are informed of the resource allocation through training signaling. This enables estimation of both the precoded channel $\vect{H}_k \vect{W}_{k}$ and the second-order interference term $\boldsymbol{\mathcal{I}}_k=\sum_{\ell \neq k} \vect{H}_k \vect{W}_{\ell} \vect{W}_{\ell}^H \vect{H}_k^H$, both being necessary for coherent reception.
As a nice by-product \cite{Trivellato2008a}, this enables user $k$ to replace $\widetilde{\vect{C}}_k$ with the rate-maximizing MMSE receive combiner $\vect{C}_k^{\text{MMSE}}$ containing the $d_k$ dominating left singular vectors of $( \vect{I}_{M} + \boldsymbol{\mathcal{I}}_k )^{-1} \vect{H}_k \vect{W}_{k}$ \cite{Guthy2010a}.
This improves the information rate by balancing between signal gain and interference rejection. We consider $\widetilde{\vect{C}}_k$ in the analysis, while $\vect{C}_k^{\text{MMSE}}$ is used in simulations.
\end{remark}

\subsection{Linear Precoding: BD and ZFC}
\label{subsection_definition_BD_ZF}

In this paper, we primarily analyze and compare two instances of \eqref{eq_practical_resource_allocation}: \emph{block-diagonalization (BD)} \cite{Spencer2004a} and \emph{zero-forcing with combining (ZFC)} \cite{Jindal2008a,Trivellato2008a}. These strategies allocate a fixed number of streams per scheduled user, but can be combined with any scheduling policy. There are alternative strategies that allocate different numbers of streams to different users \cite{Boccardi2007a}, but simulations will show that these are \emph{not} increasing the performance when the CSI acquisition overhead is treated properly.

\begin{definition} \emph{(Block-Diagonalization Precoding)} \label{definition_block_diagonalization}
Let $\mathcal{S}^{\text{BD}}$ be a scheduling set with at most $\frac{N}{M}$ users. For each user $k \in \mathcal{S}^{\text{BD}}$, we set $d_k=M$ and $\vect{W}_k=\vect{W}_k^{\text{BD}} \vect{\Upsilon}_k^{1/2}$, where
$\vect{W}_k^{\text{BD}}$ is a semi-unitary matrix that satisfies $\vect{W}_k^{\text{BD},H} \vect{W}_k^{\text{BD}}= \vect{I}_{M}$ and $\vect{H}_{\ell} \vect{W}_k^{\text{BD}} = \vect{0}$ for all $\ell \in \mathcal{S}^{\text{BD}} \! \setminus \!\{k\}$. The power allocation is given by the diagonal matrix $\vect{\Upsilon}_k \succeq \vect{0}_M$.
The information rate is
\begin{equation} \label{eq_data_rate_BD}
g^{\text{BD}}_k(P) = \log_2 \det \left( \vect{I}_{M} +  \vect{H}_k \vect{W}_k^{\text{BD}} \vect{\Upsilon}_k \vect{W}_k^{\text{BD},H} \vect{H}_k^H \right).
\end{equation}
\end{definition}

\begin{definition} \emph{(Zero-Forcing Precoding with Combining)} \label{definition_zero-forcing}
Each user combines its antennas using some channel-dependent unit-norm vector $\tilde{\vect{c}}_k \in \mathbb{C}^{M \times 1}$.
Based on the effective channels $\vect{h}_k^H =\tilde{\vect{c}}_k^H \vect{H}_k \in \mathbb{C}^{1 \times N}$, a scheduling set $\mathcal{S}^{\text{ZFC}}$ with at most $N$ users is selected.
For each user $k \in \mathcal{S}^{\text{ZFC}}$, we set $d_k=1$ and let $\vect{W}_k=\sqrt{p_k} \vect{w}^{\text{ZFC}}_{k}$, where $\vect{w}^{\text{ZFC}}_{k}$ is a unit-norm vector that satisfies $\vect{h}_{\ell}^H \vect{w}^{\text{ZFC}}_{k} = 0$ for all $\ell \in \mathcal{S}^{\text{ZFC}} \!\setminus\!\{k\}$. The power $p_k\geq 0$ is allocated to user $k$ and the information rate is
\begin{equation} \label{eq_data_rate_ZF}
g^{\text{ZFC}}_k(P) = \log_2 \left( 1 +  p_k | \vect{h}_k^H \vect{w}^{\text{ZFC}}_{k}|^2  \right).
\end{equation}
\end{definition}

The sum-rate maximizing power allocations for BD and ZFC are achieved through water-filling (see \cite{Spencer2004a}), but the asymptotic analysis in this paper often assumes equal power allocation (i.e., $\vect{\Upsilon}_k = \frac{P}{M |\mathcal{S}^{\text{BD}}|} \vect{I}_M \,\, \forall k \in \mathcal{S}^{\text{BD}}$ and $p_k=\frac{P}{|\mathcal{S}^{\text{ZFC}}|} \,\, \forall k \in \mathcal{S}^{\text{ZFC}}$) since this becomes optimal in the high-SNR regime where $P \rightarrow \infty$ \cite{Lee2007a}. Although the definitions of BD and ZFC assume perfect CSI, both strategies can be applied when the transmitter has imperfect CSI by making $\vect{W}_k$ orthogonal to the acquired co-user channels \cite{Jindal2008a,Trivellato2008a,Ravindran2008a}. The resulting loss will be quantified in later sections.

ZFC can schedule up to $N$ users and sends one data stream per user, while BD can only schedule $\frac{N}{M}$ users but  multiplexes $M$ streams to each of them. Although BD and ZFC are identical when each user only has one antenna, this does \emph{not} mean that BD is a generalization of ZFC. In fact, there are good reasons for applying ZFC instead of BD when $M>1$:

\begin{enumerate}
\item The base station only needs to acquire the effective channels $\vect{h}_k$;

\item The effective channel $\vect{h}_k$ has better properties than $\vect{H}_k$ and can be adapted for interference rejection;

\item User devices require simpler hardware that only decodes one stream.
\end{enumerate}

The interference mitigation is, on the other hand, less restrictive under BD since fewer users are involved and the mutual interference between streams sent to the same user is handled by receive processing \cite{Ravindran2008a}. By analyzing and comparing ZFC and BD under both perfect and imperfect CSI, we try to answer the fundamental question: should we select many multi-antenna users to enable receive combining or select few users and exploit multi-stream multiplexing?

\begin{remark}[Ambiguous Terminology]
The terminology \emph{block-diagonalization} and \emph{zero-forcing} have been given different meanings in prior works.
Herein, BD refers to the original work in \cite{Spencer2004a}, where each active user receives exactly $M$ data streams. Apart from the ZFC strategy in Definition \ref{definition_zero-forcing} (and in \cite{Jindal2008a,Trivellato2008a}), another downlink zero-forcing strategy for multi-antenna users was proposed in \cite{Yoo2006a}. In their definition, each antenna at the multi-antenna users is viewed as a separate virtual single-antenna user and the zero-forcing idea is applied to send a separate stream to each antenna
with zero inter-antenna interference. That approach is nothing else than BD with stricter interference mitigation and can \emph{never} perform better than BD.
Herein, ZFC means sending one stream per user and utilizing receive combining, thus ZFC is \emph{not} a special case of BD and can hypothetically outperform BD.
\end{remark}

\section{Comparison of BD and ZFC with Perfect CSI}
\label{section_comp_BD_ZFC_perfect_CSI}

In this section, we will compare BD and ZFC in the ideal scenario when both the base station and the users have perfect CSI. We derive analytic results indicating the impact of different system properties. Under perfect CSI,  the achievable sum rate in \eqref{eq_practical_resource_allocation} asymptotically becomes (as $P \rightarrow \infty$) \cite{Lee2007a}
\begin{equation} \label{eq_sum_rate_asymptotic}
\begin{split}
f_{\text{sum}}^{\text{BD}}(P) &\cong N \log_2 \! \left(\frac{P}{N}\right) \!+ \!\! \sum_{k \in \mathcal{S}^{\text{BD}}} \! \log_2 \det (\vect{H}_k \vect{W}_k^{\text{BD}} \vect{W}_k^{\text{BD},H} \vect{H}_k^H), \\
f_{\text{sum}}^{\text{ZFC}}(P) &\cong N \log_2 \! \left(\frac{P}{N}\right) \!+ \!\!\sum_{k \in \mathcal{S}^{\text{ZFC}}} \! \log_2 (|\vect{h}^H_k \vect{w}^{\text{ZFC}}_{k}|^2),
\end{split}
\end{equation}
for BD and ZFC, respectively. This result is based on having scheduling sets that satisfy $|\mathcal{S}^{\text{BD}}|=\frac{N}{M}$ and $|\mathcal{S}^{\text{ZFC}}|=N$ and on equal power allocation (which is asymptotically optimal).

For both strategies, the asymptotic sum rate behaves as $\mathcal{M}_{\infty} \log_2(P) + \mathcal{R}_{\infty}$, where $\mathcal{M}_{\infty}$ is the multiplexing gain and $\mathcal{R}_{\infty}$ is the rate offset. Both BD and ZFC achieve a multiplexing gain of $\mathcal{M}_{\infty}=N$, which is the same high-SNR slope as of the sum capacity. We thus need to compare the rate offsets $\mathcal{R}_{\infty}$ to conclude which strategy is preferable in the high-SNR regime.

\begin{theorem} \label{theorem_asymptotic_difference}
Assume the receive correlation matrices $\vect{R}_{R,k}$ have eigenvalues $\lambda_{k,M} \geq \ldots \geq \lambda_{k,1} > 0$ and the use of random user selection with $|\mathcal{S}^{\text{BD}}|=\frac{N}{M}$, $|\mathcal{S}^{\text{ZFC}}|=N$. The expected asymptotic difference in sum rate between BD and ZFC (with MRC) is

\begin{equation} \label{eq_asymptotic_difference}
\begin{split}
&\bar{\beta}_{\text{BD-ZFC}} = \mathbb{E} \left\{ \lim_{P \rightarrow \infty}  f_{\text{sum}}^{\text{BD}}(P) - f_{\text{sum}}^{\text{ZFC}}(P) \right\} \\
&= N \frac{\log_2(e)}{M} \sum_{i=1}^{M-1} \frac{M-i}{i} \!+\! \log_2 \left(\prod_{k \in \mathcal{S}^{\text{BD}}} \prod_{m=1}^{M} \lambda_{k,m} \! \right) \!-\! \sum_{\ell \in \mathcal{S}^{\text{ZFC}}}  z_{\ell}
\end{split}
\end{equation}
where $z_{\ell} = \mathbb{E}\{ \log_2( \| \tilde{\vect{c}}_{\ell}^H \vect{H}_{\ell} \|_2^2) \} - \frac{\psi(N)}{\log_e(2)}$ and $\psi(\cdot)$ is the digamma function. Furthermore, $\log_2(\lambda_{\ell,M}) \leq z_{\ell} \leq \log_2( \mathbb{E}\{ \| \tilde{\vect{c}}_{\ell}^H \vect{H}_{\ell} \|_2^2 \}) - \frac{\psi(N)}{\log_e(2)}$ where $\mathbb{E}\{ \| \tilde{\vect{c}}_{\ell}^H \vect{H}_{\ell} \|_2^2  \}$ is given by \eqref{eq_mean_value_effective_channel} in Lemma \ref{lemma_distribution_effective_channel}.
\end{theorem}
\begin{IEEEproof}
The proof is given in Appendix \ref{app_theorem_asymptotic_difference}.
\end{IEEEproof}

The expected asymptotic difference in \eqref{eq_asymptotic_difference} has several terms. The first term is the (positive) expected gain of BD in a spatially uncorrelated scenario with homogenous user channels and no receive combining---this was considered in \cite[Theorem 3]{Lee2007a}. The other terms depend on the spatial correlation and choice of receive combining. For users with homogenous channel conditions where all $\vect{R}_{R,k}$ have the same eigenvalues $\lambda_{k,m}=\lambda_m$, we have
\begin{equation}
\bar{\beta}_{\text{BD-ZFC}} \leq  N \frac{\log_2(e)}{M} \sum_{i=1}^{M-1} \frac{M-i}{i} + N \log_2 \frac{\prod_{m=1}^{M} \lambda_{m}^{1/M}}{\lambda_{M}}
\end{equation}
where the last term contains the geometric mean of all eigenvalues divided by the largest eigenvalue. This ratio is smaller than one (or equal for uncorrelated channels) and thus its logarithm is negative and approaches $-\infty$ as the eigenvalue spread increases. Therefore, Theorem \ref{theorem_asymptotic_difference} shows that BD \emph{might} have an advantage on uncorrelated channels, but ZFC always becomes the better choice as the receive-side correlation grows. The explanation is that BD has less restrictive interference mitigation, but is more vulnerable to poor channels since it uses all channel dimensions for transmission. We can expect a similar impact of any channel property that increases the eigenvalue spread in $\vect{H}_k \vect{H}_k^H$; for example, spatial correlation at the transmitter-side or a strong (low-rank) line-of-sight component.

To illustrate the opposite effect of having users with different path losses, we assume for simplicity that there are $\frac{N}{M}$ strong users with $\vect{R}_{R,k} = \gamma \vect{I}_M$, for some $\gamma> 1$, and $N-\frac{N}{M}$ weak users with $\vect{R}_{R,k} = \vect{I}_M$. If BD only serves the strong users while ZFC serves also the weak users, we have
\begin{equation}
\bar{\beta}_{\text{BD-ZFC}} \leq  N \frac{\log_2(e)}{M} \sum_{i=1}^{M-1} \frac{M-i}{i} + \left( N-\frac{N}{M} \right) \log_2( \gamma).
\end{equation}
This upper bound approaches $+\infty$ as the difference $\gamma$ between the strong and weak users grows. Although not strictly proved, this indicates that BD is better at utilizing heterogenous channel conditions as it requires fewer users to be close to the base station to achieve high sum rates. This benefit reduces if some fairness mechanism is used to compensate for unfavorable path losses.

The expected asymptotic difference in sum rate, $\bar{\beta}_{\text{BD-ZFC}}$, can be transformed into a difference $-\frac{\bar{\beta}_{\text{BD-ZFC}}}{ 10 N \log_{10}(2)}$ [dB] in transmit power to achieve the same sum rate in the high-SNR regime \cite{Lee2007a}.

\subsection{Impact of User Selection}

The comparison in Theorem \ref{theorem_asymptotic_difference} was based on random user selection of the maximal number of users ($\frac{N}{M}$ with BD and $N$ with ZFC), although scheduling of spatially separated users is necessary to achieve the full potential of multi-user MIMO. This paper assumes $K \geq N$ users, meaning that only a subset of users is scheduled at each channel use. If the users are unevenly distributed in the cell, it could be beneficial to intentionally schedule fewer users than possible. We will now analyze how the ability of selecting users with spatially compatible channels impacts performance.

In the high-SNR regime, the optimal (semi-unitary) precoding matrix $\vect{W}_k^{\textrm{su}}$ for single-user transmission matches the channel as $\widetilde{\vect{C}}_k^H\vect{H}_k \vect{W}_k^{\textrm{su}} = \widetilde{\vect{C}}_k^H \vect{H}_k$, while the precoding matrix $\vect{W}_k  \in \mathbb{C}^{N \times d_k}$ of an SDMA strategy is balanced between matching the own channel and being orthogonal to the co-user channels. The expected asymptotic performance loss of having to cancel inter-user interference is therefore
\begin{equation} \label{eq_expected_precoding_loss}
\begin{split}
\mathbb{E}\{\text{Loss} \} &= \mathbb{E}\{  \log_2 \det (\widetilde{\vect{C}}_k^H \vect{H}_k \vect{H}_k^H \widetilde{\vect{C}}_k) \\ & \qquad - \log_2 \det (\widetilde{\vect{C}}_k^H \vect{H}_k \vect{W}_k \vect{W}_k^{H} \vect{H}_k^H \widetilde{\vect{C}}_k) \} \\
& = \mathbb{E}\Big\{  \log_2 \frac{\det ( \vect{\Lambda}_{k} \vect{\Lambda}_{k}^H )}{ \det ( \vect{\Lambda}_{k} \vect{B}_k \vect{W}_k \vect{W}_k \vect{B}_k^H \vect{\Lambda}_{k}^H ) } \Big\} \\
&= - \mathbb{E}\{ \log_2 \det (\vect{B}_k \vect{W}_k \vect{W}_k \vect{B}_k^H) \}
\end{split}
\end{equation}
where $\vect{\Lambda}_{k} \in \mathbb{C}^{d_k \times d_k}$ contains the non-zero singular values of $\widetilde{\vect{C}}_k^H \vect{H}_k$ and $\vect{B}_k$ contains the corresponding right singular vectors.\footnote{These matrices can be obtained from a \emph{compact} singular value decomposition $\widetilde{\vect{C}}_k^H \vect{H}_k = \vect{U}_k \vect{\Lambda}_{k} \vect{B}_k$. Note that $\vect{B}_k$ contains an orthonormal basis of the row space of the effective channel $\widetilde{\vect{C}}_k^H \vect{H}_k$.} Observe that the eigenvalues of $\vect{B}_k \vect{W}_k \vect{W}_k \vect{B}_k^H$ are smaller or equal to one, thus $\mathbb{E}\{\text{Loss} \} \geq 0$. The following theorem indicates how this loss is affected by user selection.

\begin{theorem} \label{theorem_impact_of_scheduling}
For any given scheduling sets $\mathcal{S}^{\text{BD}},\mathcal{S}^{\text{ZFC}}$ (with $|\mathcal{S}^{\text{BD}}|=\frac{N}{M}$ and $|\mathcal{S}^{\text{ZFC}}|=N$), suppose we replace one of the users in each set with the best one among $K$ random users.
If the best user is the one minimizing the expected asymptotic loss in \eqref{eq_expected_precoding_loss}, these losses for BD and ZFC, respectively, can be lower bounded as \vskip-4mm
\begin{equation}
\begin{split}
\mathbb{E}\{\text{Loss}_{\text{BD}}\} &\geq - M \log_2(1-c_1 K^{-\frac{1}{M(N-M)}}) \\
\mathbb{E}\{\text{Loss}_{\text{ZFC}} \} &\geq - \log_2(1-c_2 K^{-\frac{1}{N-M}})
\end{split}
\end{equation}
when $K$ is large ($c_1,c_2$ are positive constants, see the proof).
\end{theorem}
\begin{IEEEproof}
The proof is given in Appendix \ref{app_theorem_impact_of_scheduling}.
\end{IEEEproof}

The lower bounds in this theorem indicate that it is easier to find users with near-orthogonal channels under ZFC than under BD. This seems reasonable since the random channels of BD users occupy $M$ dimensions and should happen to be compatible to the co-users in all of them, while ZFC users only utilize one dimension and use receive combining to pick the most compatible among its $M$ dimensions. Related observations can be made in the area of channel quantization, where fewer codewords are necessary to describe $(N \times 1)$-dimensional channels to a certain accuracy than are needed for $(N \times M)$-dimensional channels \cite{Dai2008a}. The concave structure of the information rates makes it difficult to obtain exact results, but the indications of Theorem \ref{theorem_impact_of_scheduling} are verified by simulations herein.

\subsection{Numerical Illustrations under Perfect CSI}

\label{subsection_numerical_perfect_CSI}

Next, the analytic properties in Theorem \ref{theorem_asymptotic_difference} and Theorem \ref{theorem_impact_of_scheduling} are illustrated numerically.
To this end, we adopt the simple exponential correlation model of \cite{Loyka2001a}, where $0 \leq \rho \leq 1$, $\iota=\sqrt{-1}$, $U[\cdot,\cdot)$ denotes a uniform distribution, and
\begin{equation} \label{ch2_eq_exponential_correlation_model}
[\vect{R}(\rho,\theta)]_{ij} = \begin{cases} (\rho e^{\iota \theta})^{j-i}, & i \leq j, \\[-2mm]
(\rho e^{-\iota \theta})^{i-j}, & i>j,
\end{cases} \quad  \theta \sim U[0,2\pi).
\end{equation}
The magnitude $\rho$ is the \emph{correlation factor} between adjacent antennas, where $\rho=0$ means no spatial correlation and $\rho=1$ means full correlation. For simplicity, $\rho$ is the same for all users while $\theta$ is different. Note that $\rho$ impacts the perceived spatial correlation non-linearly; a typical angular spread in a highly spatially correlated scenario is $10-20$ degrees which roughly corresponds to $\rho \approx 0.9$ \cite{Bjornson2009c}.

\begin{figure}
\begin{center}
\includegraphics[width=86mm]{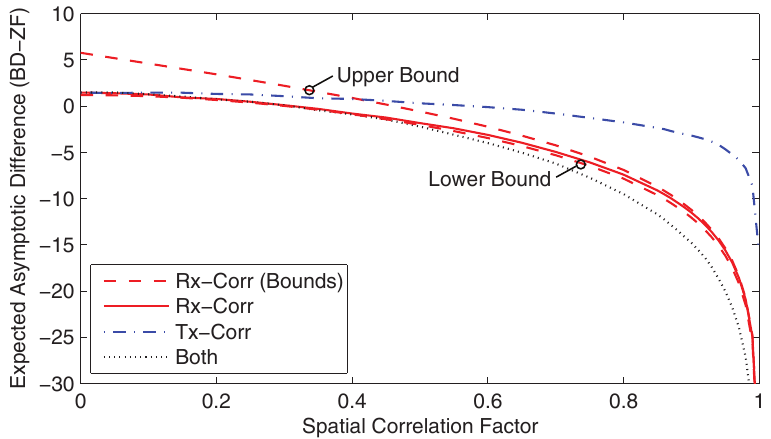} \vskip-2mm
\caption{The expected asymptotic difference between BD and ZFC in a system with $N=8$ transmit antennas, $M=2$ receive antennas per user, and random user selection.
The impact of spatial correlation at the receiving users, transmitting base station, and both sides is shown (using the exponential correlation model from \cite{Loyka2001a} with different correlation factors $\rho$).}\label{figure_correlation}
\end{center} \vskip-3mm
\end{figure}

The expected asymptotic difference between BD and ZFC is shown in Fig.~\ref{figure_correlation} as a function of $\rho$, using $N=8$ transmit antennas and $M=2$ receive antennas. This simulation confirms that BD is advantageous in uncorrelated systems, while ZFC becomes beneficial as the correlation increases ($\rho>0.4$ under receive-side correlation, $\rho>0.7$ under transmit-side correlation, and $\rho>0.25$ when both sides are correlated). The two bounds from Theorem \ref{theorem_asymptotic_difference} are also shown in the Fig.~\ref{figure_correlation}. The lower bound is very accurate, while the upper bound is only tight at high correlation.

\begin{figure}
\begin{center}
\includegraphics[width=86mm]{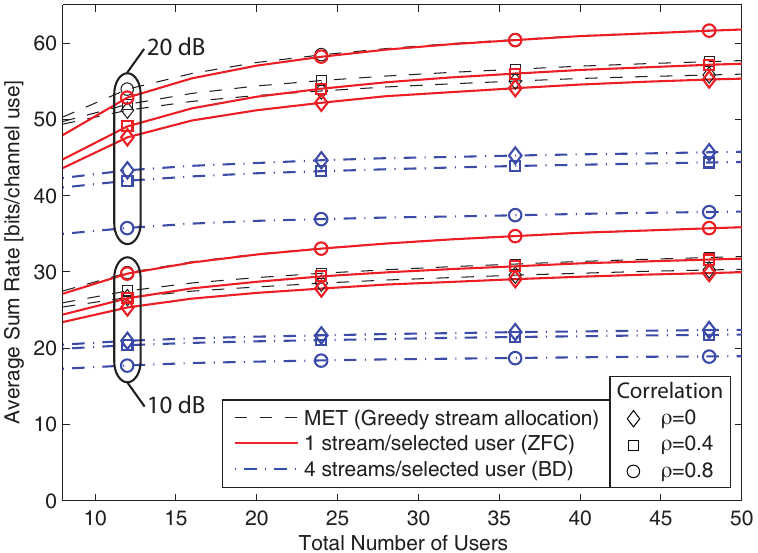} \vskip-2mm
\caption{The average achievable sum rate in a system with perfect CSI, $N=8$ transmit antennas, $M=4$ receive antennas, and the same average SNR among all users (10 or 20 dB).
The performance with different strategies are shown as a function of the total number of users and for different correlation factors $\rho$ among the receive antennas.}\label{figure_perfect_dBs}
\end{center} \vskip-5mm
\end{figure}

To exemplify the impact of user selection, we use the \emph{capacity-based suboptimal user selection} (CBSUS) algorithm from \cite{Shen2006a}, which greedily adds users sequentially to maximize the sum rate and might give scheduling sets with fewer than $N$ data streams. We consider a scenario with $N=8$ uncorrelated transmit antennas and $M=4$ receive antennas with correlation factor $\rho \in \{0,0.4,0.8\}$; see \cite{Bjornson2011e} for another scenario. We compare ZFC (1 stream/user) and BD (4 streams/user) with multi-user eigenmode transmission (MET) from \cite{Boccardi2007a} where data streams are allocated greedily with zero inter-user interference and users can have different numbers of streams. We also simulated 2 streams/user, but it is not shown herein because the sum rate was always in between ZFC and BD.

Fig.~\ref{figure_perfect_dBs} shows the average achievable sum rate as a function of the total number of users $K$. We consider the case when all users have the same average SNR (defined as $P \frac{ \mathbb{E} \{\| \vect{H}_k\|_F^2 \}}{NM}$), either equal to 10 or 20 dB. Irrespective of the SNR, number of users, and receive-side correlation, ZFC outperforms BD. Thus, the scheduling-benefit of ZFC (from Theorem \ref{theorem_impact_of_scheduling}) dominates over the interference mitigation-benefit of BD (from Theorem \ref{theorem_asymptotic_difference})---even for spatially uncorrelated channels. As expected, the performance with ZFC improves with $\rho$, while correlation degrades the BD performance. MET has an advantages over ZFC since it can allocate different numbers of streams to different users (based on how many singular values are strong in their channels), but this advantage is small and disappears asymptotically with the number of users; this was also observed in \cite{Boccardi2007a}.

\begin{figure}
\begin{center}
\includegraphics[width=86mm]{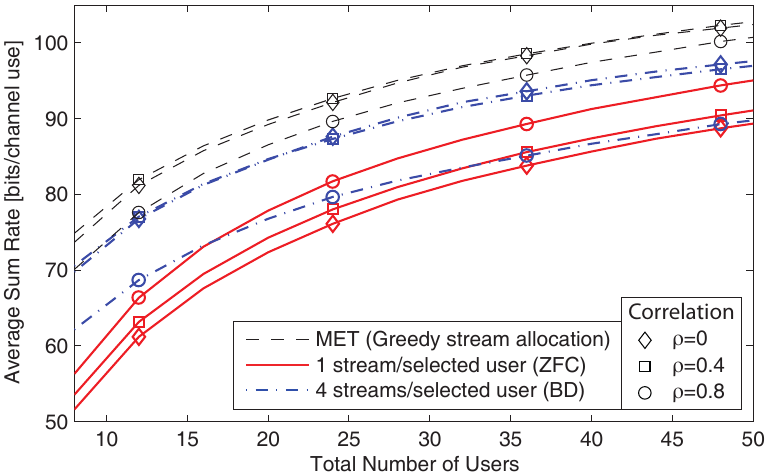} \vskip-2mm
\caption{The average achievable sum rate in a circular cell with perfect CSI, $N=8$ transmit antennas, $M=4$ receive antennas, and an SNR of 20 dB at the cell edge.
The performance with different strategies are shown as a function of the total number of users and for different correlation factors $\rho$ among the receive antennas.}\label{figure_perfect_pathloss}
\end{center} \vskip-5mm
\end{figure}

\begin{figure}
\begin{center}
\includegraphics[width=86mm]{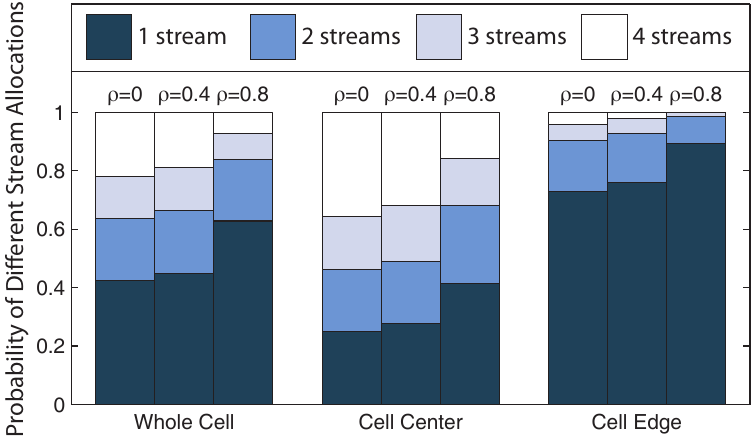} \vskip-2mm
\caption{The probability that a scheduled user is allocated a certain number of streams, assuming a circular cell with perfect CSI, $N=8$ transmit antennas, $M=4$ receive antennas, $K=20$ users, and an SNR of 20 dB at the cell edge. The whole cell has a radius of 250 meters, whereof users closer than 100 meters belong to the cell center and users further away than 200 meters belong to the cell edge.}\label{figure_bars}
\end{center} \vskip-5mm
\end{figure}

Next, we consider heterogeneous channel conditions by having uniformly distributed users in a circular cell with radius $250$ m (minimal distance is $35$ m), a path loss coefficient of 3.5, and log-normal shadow-fading with 8 dB in standard deviation. The average achievable sum rate is shown in Fig.~\ref{figure_perfect_pathloss} with an SNR of 20 dB at the cell edge.\footnote{Such SNRs are reasonable in dense cellular systems and are necessary to compare BD and ZFC in regimes where these are supposed to work well.} The variation in path loss between users makes the results very different from the previous scenario in Fig.~\ref{figure_perfect_dBs}. At low receive-correlation, BD outperforms ZFC, but the difference reduces with $K$. ZFC is however better than BD at high correlation and many users. MET has a large advantage over the other strategies, explained by its flexible stream allocation. To comprehend the difference, the probability that a selected user is allocated a certain number of streams is shown in Fig.~\ref{figure_bars}.
We observe that spatial correlation reduces the number of streams per user, but the distance-dependence is even more significant; cell center users usually receive many streams while cell edge users only receive one or a few streams. This is natural since cell center users are more probable to have channel matrices with multiple relatively strong singular directions.

The conclusion is that ZFC is the method of choice in multi-user MIMO systems with perfect CSI and homogenous user conditions (since it performs very closely to the more complicated MET). On the other hand, MET and BD are better under heterogeneous user conditions. It is worth noting that the more streams allocated per user, the more channel dimensions need to be know at the base station. The next section will therefore study how practical CSI acquisition affects our results.

\section{Comparison of BD and ZFC with Imperfect CSI}

In this section, we continue the comparison of BD and ZFC by introducing imperfect CSI, originating from either quantized feedback in an FDD system or imperfect reverse-link estimation in a TDD system. The resources for channel acquisition are limited which has a major impact on both the number of channel dimensions that can be acquired per user and the accuracy of the acquired CSI. Theoretically,
users can feed back different numbers of channel dimensions depending on some kind of long-term statistical CSI, but that would reduce the coverage (by favoring cell center users)
and require a flexible system operation with additional control signaling. We therefore assume that the system acquires $d$ dimensions/user from a randomly selected user set, where $d \geq 1$ is fixed but depends on the intended precoding strategy. This assumption is relaxed in the numerical evaluation.

\subsection{Comparison with Quantized CSI}

In the FDD system operation of Fig.~\ref{figure_fdd_operation}, each user selected for feedback conveys the $d$-dimensional subspace spanned by its effective channel $\widetilde{\vect{C}}_k^H \vect{H}_k$ using $B$ bits. Similar to \cite{Love2003a,Dai2008a,Raghavan2007a,Love2008a,Ravindran2008a}, we use a codebook $\mathcal{C}_{N,d,B}=\{\vect{U}_1,\ldots,\vect{U}_{2^{B}}\}$ with codewords
$\vect{U}_{i} \in \mathbb{C}^{N \times d}$ from the (complex) \emph{Grassmannian manifold} $\mathcal{G}_{N,d}$; that is, the set of all  $d$-dimensional linear subspaces (passing through the origin) in an $N$-dimensional space. Each codeword forms an orthonormal basis, thus $\vect{U}_{i}$ is a semi-unitary matrix satisfying $\vect{U}_{i}^H \vect{U}_{i} = \vect{I}_{d}$.
User $k$ selects the codeword that minimizes the chordal distance \cite{Zhou2006a}:
\begin{equation}
\bar{\vect{H}}_k = \argmin{ \vect{U} \in \mathcal{C}_{N,d,B}} \,\, \delta \left(  \widetilde{\vect{C}}_k^H \vect{H}_k, \vect{U} \right)
\end{equation}
where $\delta(\vect{B},\vect{U}) = \sqrt{d-\tr( \spa(\vect{B})^H \vect{U} \vect{U}^H \spa(\vect{B}) )}$ and $\spa(\cdot)$ gives a matrix containing an orthonormal basis of the row space.
We assume error-free and delay-free feedback, but the conclusions of this section are expected to hold true also under feedback errors (cf.~\cite{Caire2010a}).

There is a variety of ways to handle feedback errors (especially if the error structure is known), but a simple approach is to treat $\bar{\vect{H}}_k$ as being the true channel \cite{Ravindran2008a} and calculate the precoding using a strategy developed for perfect CSI. This results in a lower bound on the performance and the information rates with BD and ZFC becomes
\begin{align} \label{eq_data_rate_BD_quant}
\!\! g^{\text{BD-Q}}_k(P) &= \log_2 \! \frac{\det \!\Big( \vect{I}_{M} \!+\!\!\! \fracSum{\ell \in \mathcal{S}^{\text{BD}}} \vect{H}_k \bar{\vect{W}}_{\ell}^{\text{BD}} \bar{\vect{\Upsilon}}_{\ell} \bar{\vect{W}}_{\ell}^{\text{BD},H} \vect{H}_k^H \! \Big) }{ \! \det \! \Big(\vect{I}_{M} \!+\!\!\! \fracSum{\ell \in \mathcal{S}^{\text{BD}} \setminus \{k\}} \!\!\! \vect{H}_k \bar{\vect{W}}_{\ell}^{\text{BD}} \bar{\vect{\Upsilon}}_{\ell} \bar{\vect{W}}_{\ell}^{\text{BD},H} \vect{H}_k^H \!\Big) \!\! } \\
\label{eq_data_rate_ZF_quant}
\!\! g^{\text{ZFC-Q}}_k(P) &= \log_2 \! \bigg( \!1 \!+\!  \frac{ \bar{p}_k |\vect{c}_k^H \vect{H}_k \bar{\vect{w}}^{\text{ZFC}}_{k}|^2 }{1 + \!\!\fracSum{\ell \in \mathcal{S}^{\text{ZFC}} \setminus \{k\}} \!\! \bar{p}_{\ell} |\vect{c}_k^H \vect{H}_k \bar{\vect{w}}^{\text{ZFC}}_{\ell}|^2   }  \! \bigg)
\end{align}
for users in the scheduling sets $\mathcal{S}^{\text{BD}}$ and $\mathcal{S}^{\text{ZFC}}$, respectively.

Next, we quantify the performance loss for BD and ZFC compared with having perfect CSI.
Random vector quantization (RVQ) is used for analytic convenience (as in \cite{Santipach2004a,Au-Yeung2007a,Jindal2008a,Ravindran2008a}), meaning that
we average over codebooks with random codewords from the Grassmannian manifold. As any judicious codebook design is better than RVQ, the upper bounds on the performance loss that we will derive are valid for any reasonable codebook. The following theorem provides an upper bound on the performance loss under BD and extends results in \cite{Ravindran2008a} to include heterogeneous user conditions and spatial correlation.

\begin{theorem} \label{theorem_rateloss_BD}
Assume that $\frac{N}{M}$ users are scheduled randomly. The average rate loss with BD (using equal power allocation) for user $k \in \mathcal{S}^{\text{BD}}$ due to RVQ is upper bounded as
\begin{equation} \label{eq_bounding_QBD_loss_theorem}
\begin{split}
\Delta^{\text{BD-Q}}_k &= \mathbb{E} \{ g^{\text{BD}}_k(P)- g^{\text{BD-Q}}_k(P) \} \\
& \leq \log_2 \det \left(\vect{I}_M + \frac{P}{M} D^{\text{BD}} \vect{R}_{R,k} \right)
\end{split}
\end{equation}
where the average quantization distortion is
\begin{equation}
\begin{split}
&D^{\text{BD}} = \mathbb{E} \{ \delta^2(\vect{H}_k,\bar{\vect{H}}_k) \} \\
&\approx \frac{\Gamma\left(\frac{1}{M(N-M)}\right)}{M(N-M)} \! \left( \!\frac{2^{B}}{(M(N-M))!} \prod_{i=1}^{M} \frac{(N-i)!}{(M-i)!} \! \right)^{\!-\frac{1}{M(N-M)}}.
\end{split}
\end{equation}
\end{theorem}
\begin{IEEEproof}
The proof is given in Appendix \ref{app_theorem_rateloss_BD}.
\end{IEEEproof}

This theorem will be compared with the corresponding result for ZFC, but before stating that result we discuss how to select the (preliminary) receive combiner $\tilde{\vect{c}}_k$.
There are primarily two factors to consider when selecting $\tilde{\vect{c}}_k$: the gain of the effective channel $\|\tilde{\vect{c}}_k^H \vect{H}_k\|_2^2$ and the quantization distortion.
The results of \cite{Kountouris2006a,Ravindran2008b} indicate that the top priority in multi-user MIMO systems is to achieve small quantization errors, because it is a prerequisite for low inter-user interference. The error can be minimized by the \emph{quantization-based combining} (QBC) approach in \cite{Jindal2008a}, where the codeword and receive combiner are selected jointly as
\begin{equation} \label{eq_QBC_receiver}
(\tilde{\vect{c}}^{\text{QBC}}_k,\bar{\vect{h}}_k) = \argmax{\substack{\vect{c}: \|\vect{c}\|_2=1 \\ \vect{u} \in \mathcal{C}_{N,1,B} }} \,\, \delta \left(\vect{H}_k^H \vect{c}, \vect{u} \right).
\end{equation}
The \emph{maximum expected SINR combiner (MESC)} in \cite{Trivellato2008a} achieves better practical performance by balancing effective channel gain and quantization distortion, but is asymptotically equal to QBC at high SNR. Since this is the regime of main interest herein, we will exploit the analytic simplicity of QBC. Observe that QBC and MESC are only used for improved feedback accuracy; the MMSE combiner in Remark \ref{remark:receive-combiner} is used to maximize the performance during transmission (this was \emph{not} done in the original QBC framework of \cite{Jindal2008a}).

The following theorem provides an upper bound on the performance loss under ZFC and extends results in \cite{Jindal2008a} to include heterogeneous user conditions and spatial correlation.

\begin{theorem} \label{theorem_rateloss_ZF}
Assume that $\vect{R}_{R,k}$ has eigenvalues $\lambda_{k,M} > \ldots > \lambda_{k,1}>0$ and that $N$ users are selected randomly.
The average rate loss for ZFC (using equal power allocation and the same $\tilde{\vect{c}}^{\text{QBC}}_k$) due to RVQ is upper bounded as
\begin{equation} \label{eq_bounding_QBC_loss_theorem}
\begin{split}
\Delta^{\text{ZFC-Q}}_k &= \mathbb{E} \{ g^{\text{ZFC}}_k(P)- g^{\text{ZFC-Q}}_k(P) \} \\ &\leq  \log_2 \left(1+ \frac{P}{N} D^{\text{QBC}} G_k \right)
\end{split}
\end{equation}
where the average quantization distortion is
\begin{equation}
D^{\text{QBC}} = \mathbb{E} \{ \delta^2(\vect{h}_k,\bar{\vect{h}}_k) \} \approx 2^{-\frac{B}{N-M}} \Big( \!\!\!
\begin{array}{c}
N-1\\[-1.2ex]
M-1
\end{array} \!\!\! \Big)^{-\frac{1}{N-M}}
\end{equation}
and the average channel gain with QBC is (where $\mu_{n} = \frac{1}{\lambda_{k,n}}$)
\begin{equation} \label{eq_L-expression}
\begin{split}
G_k =& \sum_{m=1}^{M-1} \sum_{n=1}^m \sum_{t=m+1}^{M} \frac{(N-M+1) A_{m,n,t}  }{(\mu_n\!-\!\mu_t) \condProd{i=1}{i \neq n}{m} (\mu_n\!-\!\mu_i) \condProd{j=m+1}{j \neq l}{M} \!\! (\mu_j\!-\!\mu_t)  },  \\
A_{m,n,t} =& \log_e \big(\frac{\mu_{m+1}}{\mu_{m}} \big) \frac{\mu_n^{m-1}}{(-\mu_t)^{
2+m-M}} \left(m \mu_t +\mu_n (M-m-1) \right) \\
+&\sum_{s=0}^{m-1} \sum_{r=0}^{M-m-1} \!\!
\Big( \!\!\!
\begin{array}{c}
m\\[-1.2ex]
s
\end{array} \!\!\! \Big) \!
\Big( \!\!\!
\begin{array}{c}
M-m-1\\[-1.2ex]
r
\end{array} \!\!\! \Big)
\frac{ \mu_n^{m-s} (-\mu_t)^{M-m-1-r}}{(-1)^{s} (1+s+r)} \\
\times & \left( \frac{m\!-\!s}{\mu_n} + \frac{M\!-\!m\!-\!1\!-\!r}{\mu_t} \right) (\mu_m^{r+s} -\mu_{m+1}^{r+s}).
\end{split}
\end{equation}
\end{theorem}
\begin{IEEEproof}
The proof is given in Appendix \ref{app_theorem_rateloss_ZF}.
\end{IEEEproof}

The rate loss expressions in Theorem \ref{theorem_rateloss_BD} and Theorem \ref{theorem_rateloss_ZF} for BD and ZFC, respectively, indicate the joint impact of spatial correlation (at the receiver) and CSI quantization on the performance.
The main observation is that spatial correlation only has a marginal effect on the feedback accuracy; the expressions have a similar structure as for uncorrelated channels and the same scaling
in the number of feedback bits is necessary to achieve the maximal multiplexing gain \cite{Jindal2008a,Ravindran2008a}.

\begin{corollary} \label{cor_multiplexing_gain_quantized}
To achieve the maximal multiplexing gain with BD or ZFC under quantized CSI and arbitrary receive correlation, it is sufficient to scale the total number of CDI feedback bits for the scheduled users as
\begin{equation} \label{eq_total_bits}
B_{\mathrm{total}} \approx N(N-M) \log_2(P) + \mathcal{O}(1).
\end{equation}
\end{corollary}

While this corollary only provides a sufficient condition, we can expect the scaling law in \eqref{eq_total_bits} to also be necessary.\footnote{The necessary scaling can be proved for ZFC with QBC using a technique from \cite[Theorem 4]{Jindal2006a}, while simulations in \cite{Ravindran2008a} show that quantized ZFC and BD have the same scaling in the necessary number of bits.} In any case, the scaling law in \eqref{eq_total_bits} is easily satisfied by allocating (approximately) $N-M$ channel uses for CSI feedback, since typically the uplink sum rate also behaves as $N \log_2(P) + \mathcal{O}(1)$ in the high-SNR regime \cite{Caire2010a}.

Observe that this result is based on random user selection, while additional feedback of gain information is necessary to achieve multi-user diversity or short-term rate adaptation (cf.~\cite{Yoo2007a}). As BD requires $M$ times more bits per user, ZFC can typically achieve feedback from $M$ times more users. We therefore expect ZFC to further strengthen its advantage at finding near-orthogonal users (indicated in Theorem \ref{theorem_impact_of_scheduling} under perfect CSI). In addition, spatial correlation at the transmitter-side (and other factors that make the channel matrices ill-conditioned) will inflict larger performance losses on BD than ZFC, just as in the case of perfect CSI.

\subsection{Comparison under Estimated CSI}

Next, we assume that the base station acquires CSI through imperfect CSI estimation. The primary focus will be on TDD systems, where channel estimates are obtained through training signaling in the uplink (assuming perfect channel reciprocity). It is worth noting that this approach is similar to having analog CSI feedback in FDD systems, where the unquantized channel coefficients are sent on an uplink subcarrier \cite{Ravindran2008a,Caire2010a}.\footnote{Digital/quantized feedback might be beneficial over analog/unquantized feedback when there is plenty of resources for channel estimation \cite{Caire2010a}. But if very accurate CSI is required, Corollary \ref{cor_multiplexing_gain_quantized} shows that the quantization codebooks grow very large and thus the search for the best codeword might be computationally infeasible.}

The reciprocal uplink counterpart to the system model in \eqref{eq_system_model} is
\begin{equation}
\widetilde{\vect{y}}_k = \vect{H}^T_k \widetilde{\vect{x}}_k + \widetilde{\vect{n}}_k
\end{equation}
where $\widetilde{\vect{y}}_k \in \mathbb{C}^{N \times 1}$ is the received uplink signal, $\widetilde{\vect{x}}_k \in \mathbb{C}^{M \times 1}$ is the transmitted uplink signal, and $\widetilde{\vect{n}}_k \sim \mathcal{CN}(\vect{0},\sigma^2 \vect{I}_{N})$ is the noise vector.\footnote{The downlink noise vector was normalized towards the channel matrix in the system model of \eqref{eq_system_model}. To account for a different noise level at the base station, $\sigma^2$ is the (relative) uplink noise variance.}
To estimate $\widetilde{\vect{C}}_k^H \vect{H}_k \in \mathbb{C}^{d \times N}$, user $k$ sends $\widetilde{\vect{C}}_k^* \vect{T}_k$ over $d$ uplink channel uses, for some known training matrix $\vect{T}_k \in \mathbb{C}^{d \times d}$ and where $(\cdot)^*$ denotes the complex conjugate.
Assuming perfect statistical CSI, the MMSE estimate $\widehat{\vect{H}}_k$ of $\widetilde{\vect{C}}_k^H \vect{H}_k$ and the corresponding error covariance matrix $\vect{E}_k$ are~\cite{Bjornson2010a}
\begin{equation} \label{eq_MMSE_estimator}
\begin{split}
\vectorize(\widehat{\vect{H}}_k^T) &= \frac{1}{\sigma^2} \vect{E}_k \widetilde{\vect{T}}_k^H \vectorize(\vect{Y}_k), \\
\vect{E}_k &= \left( ( \widetilde{\vect{C}}_k^H \vect{R}_{R,k} \widetilde{\vect{C}}_k \! \kron \vect{I}_N )^{-1} + \frac{\widetilde{\vect{T}}_k^H \widetilde{\vect{T}}_k}{\sigma^2}  \right)^{-1}
\end{split}
\end{equation}
where $\widetilde{\vect{T}}_k=(\vect{T}_k^T \! \kron \, \vect{I}_N)$ and $\vect{Y}_k$ is the received signal from training signaling.
The training matrix $\vect{T}_k$ has a total training power/SNR constraint $\tr( \vect{T}_k^H \vect{T}_k ) = \Psi$.

As under quantized CSI, we calculate the precoding by treating $\widehat{\vect{H}}_k$ as the true channel. This results in a lower bound on the performance and the information rates with BD and ZFC becomes
\begin{align} \label{eq_data_rate_BD_est}
g^{\text{BD-EST}}_k(P) &= \log_2 \! \frac{\det \!\Big( \vect{I}_{M} \!+\! \fracSum{\ell \in \mathcal{S}^{\text{BD}}} \vect{H}_k \widehat{\vect{W}}_{\ell}^{\text{BD}} \widehat{\vect{\Upsilon}}_{\ell} \widehat{\vect{W}}_{\ell}^{\text{BD},H} \vect{H}_k^H  \Big) }{ \! \det \! \Big( \vect{I}_{M} \!+\! \!\! \fracSum{\ell \in \mathcal{S}^{\text{BD}} \setminus \{k\}} \!\!\!\! \vect{H}_k \widehat{\vect{W}}_{\ell}^{\text{BD}} \widehat{\vect{\Upsilon}}_{\ell} \widehat{\vect{W}}_{\ell}^{\text{BD},H} \vect{H}_k^H \Big) \!\! } \\
\label{q_data_rate_ZF_est}
g^{\text{ZFC-EST}}_k(P) &= \log_2 \Bigg( 1 +  \frac{ \hat{p}_k |\vect{c}_k^H \vect{H}_k \widehat{\vect{w}}^{\text{ZFC}}_{k}|^2 }{1 + \fracSum{\ell \in \mathcal{S}^{\text{ZFC}} \setminus \{k\}} \hat{p}_{\ell} |\vect{c}_k^H \vect{H}_k \widehat{\vect{w}}^{\text{ZFC}}_{\ell}|^2   }  \Bigg)
\end{align}
for users in the scheduling sets $\mathcal{S}^{\text{BD}}$ and $\mathcal{S}^{\text{ZFC}}$, respectively.
The following theorem provides an upper bound on the performance loss under BD due to imperfect CSI estimation.

\begin{theorem} \label{theorem_rateloss_BD_EST}
Assume that $\frac{N}{M}$ users are scheduled randomly under BD. The average rate loss for user $k \in \mathcal{S}^{\text{BD}}$ (using equal power allocation) due to CSI estimation is upper bounded as
\begin{equation} \label{eq_bounding_BD_est_loss_theorem}
\begin{split}
\Delta^{\text{BD}} &= \mathbb{E} \{ g^{\text{BD}}_k(P)- g^{\text{BD-EST}}_k(P) \} \\
&\leq \log_2 \det \left(\vect{I}_M + \frac{P(N-M)}{N} \left( \vect{R}_{R,k}^{-T} + \frac{\vect{T}_k^H \vect{T}_k}{\sigma^2} \right)^{\!-1} \right)\!.
\end{split}
\end{equation}
\end{theorem}
\begin{IEEEproof}
The proof is given in Appendix \ref{app_theorem_rateloss_BD_EST}.
\end{IEEEproof}

This theorem will be compared with the corresponding result for ZFC, but before stating that theorem we need to consider the impact of having MRC as the receive combiner $\tilde{\vect{c}}_k$.
ZFC is similar to applying BD to the effective channels $\vect{h}_k^H =\tilde{\vect{c}}_k^H \vect{H}_k$, but an important difference is that the effective channels are not Rayleigh fading because $\tilde{\vect{c}}_k$ depends on the current channel realization. The expression in \eqref{eq_MMSE_estimator} will therefore not give the MMSE estimate, but fortunately the linear MMSE (LMMSE) estimator from a similar expression to \eqref{eq_MMSE_estimator} if we know the first two moments of $\vect{h}_k$ \cite{Bjornson2010a}.

\begin{lemma} \label{lemma_distribution_effective_channel}
Assume that $\vect{R}_{R}$ has eigenvalues $\lambda_{M} > \ldots > \lambda_{1}>0$, where the user indices were dropped for convenience.
If $\tilde{\vect{c}}$ is the dominating left singular vector of $\vect{H}$, it holds that
\begin{itemize}
\item the direction $\frac{\vect{h}}{\|\vect{h}\|_2}$ of $\vect{h}= \tilde{\vect{c}}^H \vect{H}$ is isotropically distributed on the unit sphere;
\item the gain $\|\vect{h}\|_2^2$ is independent of the direction and
\end{itemize}
\begin{equation} \label{eq_mean_value_effective_channel}
\begin{split}
&\mathbb{E}\{ \|\vect{h}\|_2^2 \}  = \! \! \sum_{m=1}^{M} \sum_{\boldsymbol{\zeta} \in
\mathcal{A}_M} \!\! \frac{ \condProdtwo{{\ell}=1}{M}
\lambda_{\zeta_{\ell}}^{N-{\ell}+1} \condProdtwo{{\ell}=N-M+1}{N} \!({\ell}-1)! }
{(-1)^{\mathrm{per}(\boldsymbol{\zeta})+m+1} \det (\vect{\Delta}) }
\\ & \times \sum_{\boldsymbol{\beta} \in \mathcal{B}_{m,M}} \!\!
\sum_{{\ell}=0}^{K_m(\boldsymbol{\beta})} \!\! \sum_{\tilde{k} \in
\widetilde{\Omega}_{\ell}^{(m)}} \frac{{\ell}!}{\tilde{k}_1! \cdots
\tilde{k}_m!}
\frac{(\fracSumtwo{i=1}{m}
\lambda_{\zeta_{\beta_i}}^{-1})^{-({\ell}+1)}}{\condProdtwo{i=1}{m}
\lambda_{\zeta_{\beta_i}}^{\tilde{k}_i}}
\end{split}
\end{equation}
where the $ij$th element of $\vect{\Delta} \in \mathbb{R}^{M \times M}$ is given by
\begin{equation}
[\vect{\Delta}]_{ij} = \lambda_j^{N-i+1} (N-i)!.
\end{equation}
In \eqref{eq_mean_value_effective_channel}, the set of all permutations of $\{1,\ldots,M\}$ is denoted $\mathcal{A}_M$. The sign of a given permutation $\boldsymbol{\zeta}=\{\zeta_1,\ldots,\zeta_M\} \in \mathcal{A}_M$ is denoted
$(-1)^{\mathrm{per}(\boldsymbol{\zeta})}$, where $\mathrm{per}(\cdot)$ is the number of inversions\footnote{An inversion in a sequence is a pair of numbers that is in incorrect order (i.e., not in ascending order).} in the permuted sequence.
Next, $\mathcal{B}_{l,M}$ is the collection of all subsets of $\mathcal{A}_M$ with cardinality $l$ and increasing elements (i.e., $\beta_1<\ldots<\beta_l$ for $\boldsymbol{\beta}=\{\beta_1,\ldots,\beta_l\} \in \mathcal{B}_{l,M}$).
The upper bound in the summation over $\ell$ is $K_l(\boldsymbol{\beta})= \sum_{i=1}^l (N-\beta_l)$.
Finally, $\widetilde{\Omega}^{(l)}_{{\ell}}$ is the set of all $l$-length partitions $\{\tilde{k}_1,\ldots,\tilde{k}_l\}$ of ${\ell}$ (i.e., $\sum_{i=1}^l \tilde{k}_i = \ell$)
that satisfy $0 \leq \tilde{k}_i \leq N-\beta_i$:
\begin{equation}
\widetilde{\Omega}^{(l)}_{{\ell}} =\bigg\{
\{\tilde{k}_1,\ldots,\tilde{k}_l\}: \,\, \sum_{j=1}^{\ell} \tilde{k}_j \!=
{\ell}, \, 0 \leq \tilde{k}_j \leq N-\beta_j \,\, \forall j \bigg\}.
\end{equation}
\end{lemma}
\begin{IEEEproof}
The proof is given in Appendix \ref{app_lemma_distribution_effective_channel}.
\end{IEEEproof}

The following theorem provides an upper bound on the performance loss under ZFC due to imperfect CSI estimation.

\begin{theorem} \label{theorem_rateloss_ZF_EST}
Assume that $N$ users are scheduled randomly under ZFC and that MRC is applied. The average rate loss for user $k \in \mathcal{S}^{\text{ZFC}}$ due to CSI estimation is upper bounded as
\begin{equation} \label{eq_bounding_ZF_EST_loss_theorem}
\begin{split}
\Delta^{\text{ZFC-EST}} &= \mathbb{E} \{ g^{\text{ZFC}}_k(P)- g^{\text{ZFC-EST}}_k(P) \} \\
&\leq  \log_2 \left(1 + \frac{P(N-1)}{N} \frac{1}{\mathbb{E}\{ \|\vect{h}_k\|_2^2 \}^{-1}+ \frac{\Psi}{\sigma^2}} \right)
\end{split}
\end{equation}
where $\mathbb{E}\{ \|\vect{h}_k\|_2^2 \}$ is given in \eqref{eq_mean_value_effective_channel}.
\end{theorem}
\begin{IEEEproof}
The proof is given in Appendix \ref{app_theorem_rateloss_ZF_EST}.
\end{IEEEproof}

The rate loss expressions in Theorem \ref{theorem_rateloss_BD_EST} and Theorem \ref{theorem_rateloss_ZF_EST} indicate the joint impact of spatial correlation and imperfect channel estimation on the performance of BD and ZFC, respectively. BD is slightly more resilient to CSI uncertainty, since the BD expression contains $(N-M)$ where the ZFC expression has $(N-1)$. But observe that the performance losses are calculated against the same precoding strategy with perfect CSI; we know from Section \ref{section_comp_BD_ZFC_perfect_CSI} that ZFC and BD have different preferable user conditions, making it hard to analytically conclude which strategy to use under imperfect CSI estimation. However, the important result is the following extension of \cite{Caire2010a} to spatially correlated scenarios with $M\geq1$.

\begin{corollary} \label{cor_multiplexing_gain_estimated}
To achieve the maximal multiplexing gain with BD or ZFC under imperfect CSI estimation and arbitrary receive correlation, it is necessary and sufficient to scale the training power $\Psi$ as
\begin{equation} \label{eq_training_law}
\frac{P}{\Psi} \rightarrow \texttt{constant} < \infty \quad \text{when} \, \, P \rightarrow \infty.
\end{equation}
\end{corollary}
\begin{IEEEproof}
The proof is given in Appendix \ref{app_cor_multiplexing_gain_estimated}.
\end{IEEEproof}

This corollary says that the training power/SNR should increase linearly with the transmit power/SNR to achieve the optimal sum rate scaling. This is, for example, satisfied by setting the total training power to $\Psi=P$ under ZFC and $\Psi=M P$ under BD, which corresponds to the reasonable assumption of having the same average SNR in the downlink and in the uplink.\footnote{Battery-powered user devices might operate at lower power budget than the base station, but Corollary \ref{cor_multiplexing_gain_estimated} is satisfied as long as $P$ and $\Psi$ exhibit the same scaling. In practical scenarios, the path loss is the main source of SNR variations and affects the downlink and uplink equally.} The demands for higher CSI accuracy with increasing SNR is therefore automatically fulfilled by the reduced estimation errors.
Observe that one uplink channel use is consumed per user antenna dimension that is estimated, thus creating a practical bound on how many user channels that can be estimated in block fading systems \cite{Caire2010a}. As ZFC only has one effective antenna per user, it can accommodate $M$ times more users than BD on the same estimation overhead and thereby exploit multi-user diversity to a larger extent.

\section{Numerical Illustrations Under Imperfect CSI}
\label{subsection_numerical_imperfect_CSI}

This section consists of two parts. First, the numerical illustrations in Section \ref{subsection_numerical_perfect_CSI} are continued under imperfect CSI estimation. Then, we analyze the performance behavior under quantized CSI.

\subsection{Continuation of Section \ref{subsection_numerical_perfect_CSI} under Estimated CSI}

\begin{figure}
\begin{center}
\includegraphics[width=86mm]{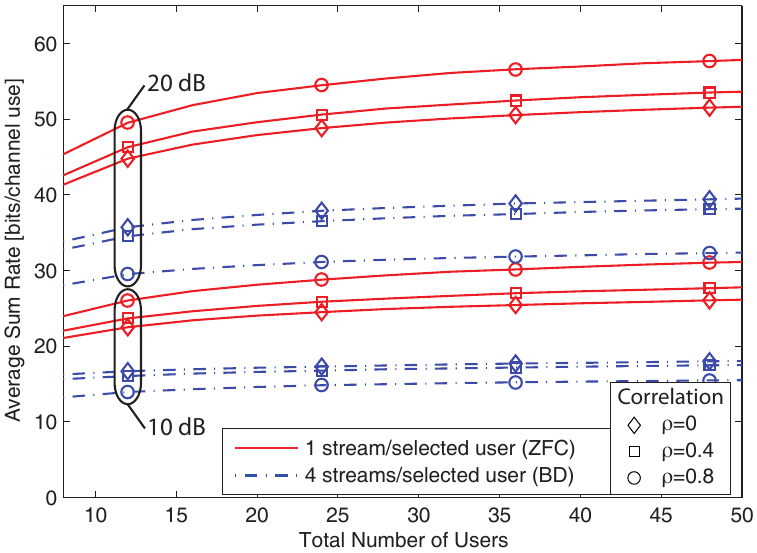} \vskip-2mm
\caption{The average achievable sum rate in a system with CSI estimation errors, $N=8$ transmit antennas, $M=4$ receive antennas, and the same average SNR among all users (10 or 20 dB).
The performance with different strategies are shown as a function of the total number of users and for different correlation factors $\rho$ among the receive antennas.}\label{figure_imperfect_dBs}
\end{center} \vskip-5mm
\end{figure}

We continue the simulations in Section \ref{subsection_numerical_perfect_CSI} by introducing imperfect CSI estimation.
We use the MSE-minimizing training matrices from \cite[Theorem 1]{Bjornson2010a} and training power $\Psi=  P \, d$ (for estimation of $d$ dimensions/user). The CBSUS algorithm in \cite{Shen2006a} is modified\footnote{Estimation errors contribute an average interference of $P (|\mathcal{S}|-1)/|\mathcal{S}| \vect{E}_{\text{est}}$, where $\vect{E}_{\text{est}} \!=\!(\vect{R}_{R,k}^{-T}\! +\! \vect{T}_k^H \vect{T}_k/\sigma^2 )^{-1}$ for BD and $\vect{E}_{\text{est}}\!=\!(1/\mathbb{E}\{\|\vect{h}_k\|_2^2\} \!+\! \Psi/\sigma^2)^{-1}$ for ZFC.} to include the average interference (due to CSI estimation errors) in the scheduling.

The average achievable sum rate is shown in Fig.~\ref{figure_imperfect_dBs} as a function of the number of users that we obtain CSI estimates for using ZFC (while BD only obtains channel estimates for $\frac{1}{M}$ of them). All users have the same average SNR of either 10 or 20 dB.
The performance loss compared with having perfect CSI is 10-20\% (see Fig.~\ref{figure_perfect_dBs}), but the conclusion is otherwise the same and even clearer than before:
ZFC outperforms BD in terms of performance with few users, in handling spatial correlation, and in exploiting multi-user diversity.

\begin{figure}
\begin{center}
\includegraphics[width=86mm]{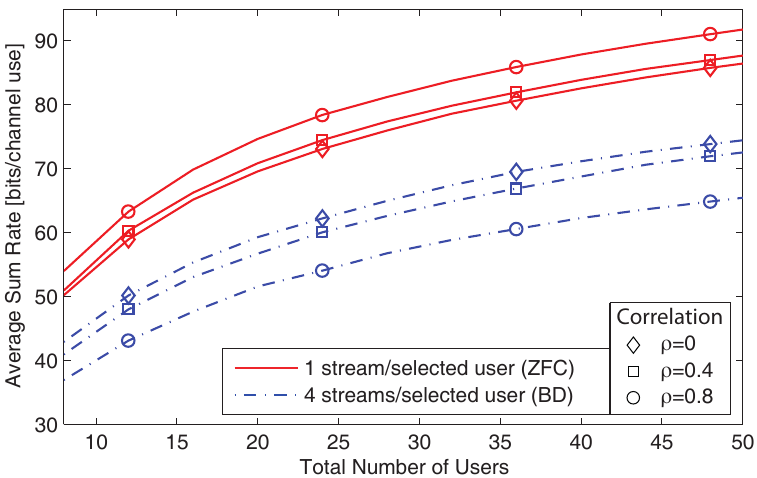} \vskip-2mm
\caption{The average achievable sum rate in a circular cell with CSI estimation errors, $N=8$ transmit antennas, $M=4$ receive antennas, and an SNR of 20 dB at the cell edge.
The performance with different strategies are shown as a function of the total number of users and for different correlation factors $\rho$ among the receive antennas.}\label{figure_imperfect_pathloss}
\end{center} \vskip-3mm
\end{figure}

\begin{figure}
\begin{center}
\includegraphics[width=86mm]{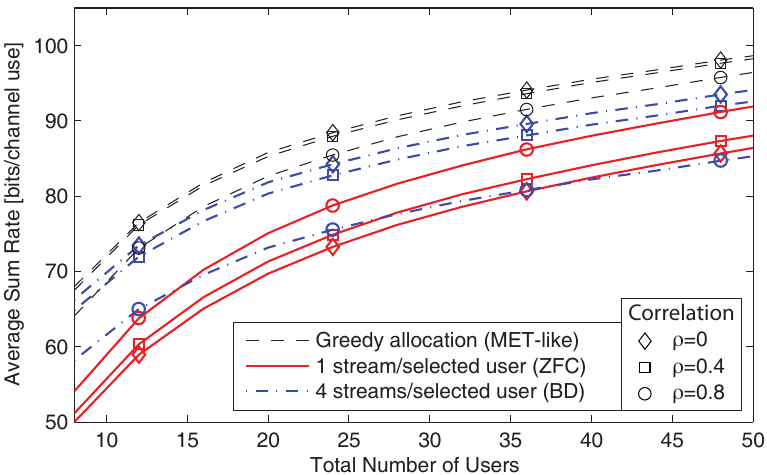} \vskip-2mm
\caption{The average achievable sum rate in a circular cell with CSI estimation errors, $N=8$ transmit antennas, $M=4$ receive antennas, and an SNR of 20 dB at the cell edge.
The performance is shown as a function of the total number of users $K$, and CSI is only acquired for the users with strongest long-term statistics.}\label{figure_imperfect_pathloss_opportunistic}
\end{center} \vskip-5mm
\end{figure}

In case of a circular cell (see Section \ref{subsection_numerical_perfect_CSI} for details), the average achievable sum rate is shown in Fig.~\ref{figure_imperfect_pathloss}. Recall from Fig.~\ref{figure_perfect_pathloss} that BD was often better than ZFC in this scenario under perfect CSI, but the case is completely different under imperfect CSI; ZFC outperforms the other strategies when the limited resources for CSI acquisition are taken into account. This means that the ZFC benefit of easily finding near-orthogonal users (among $M$ times more users than with BD) dominates the BD benefit of multi-stream multiplexing (preferably to cell center users). We also tested a MET-like strategy with greedy stream allocation (we took the optimum among feeding back 1, 2 or 4 channel dimensions per active user), but it was always identical to ZFC (in both Fig.~\ref{figure_perfect_dBs} and Fig.~\ref{figure_perfect_pathloss})---this further confirms our conclusion.

The users selected for feedback were chosen randomly (e.g., in a round-robin fashion) in Figs.~\ref{figure_imperfect_dBs} and \ref{figure_imperfect_pathloss}, but could theoretically be based on some kind of long-term statistical CSI. This could for instance mean that ZFC acquires one dimension from each of the $K$ users, while BD acquires $M$ dimensions from the $\frac{K}{M}$ users
with the strongest long-term statistics $\tr( \vect{R}_{T,k}) \tr( \vect{R}_{R,k})$. The greedy stream allocation strategy MET in \cite{Boccardi2007a} can be generalized to this scenario by finding the $K$ strongest statistical eigendirections among the users and acquire CSI for an equivalent number of dimensions per user.
Under these assumptions, the average achievable sum rate for the circular cell is shown in Fig.~\ref{figure_imperfect_pathloss_opportunistic}. The performance behavior is quite similar to the case with perfect CSI in Fig.~\ref{figure_perfect_pathloss}; BD is better than ZFC, except at high correlation, and there is a large gain from greedy stream allocation. However, we stress that this scenario is unrealistic as CSI is only acquired for cell center users, thus reducing the coverage and destroying user fairness as cell edge users are not even considered when their channels are relatively strong.

\subsection{Observations under Quantized CSI}

Next, we consider quantized CSI and let the number of feedback bits (per channel dimension) be scaled as $(N-M) \log_2(P) - \mathtt{constant}$, where the constant is selected as in \cite[Eq.~(17)]{Ravindran2008a} to maintain a 3 dB gap between BD with perfect and quantized CSI. We consider $N=4$ transmit antennas, $M=2$ receive antennas, and RVQ. We also modify\footnote{Quantization errors contribute an average interference of $P (|\mathcal{S}|-1)/|\mathcal{S}| \vect{E}_{\text{quant}}$, where $\vect{E}_{\text{quant}}=N/(M(N-M)) D^{\text{BD}} \vect{R}_{R,k}$ for BD and $\vect{E}_{\text{quant}}=D^{\text{QBC}} G /(N-1)$ for ZFC. When calculating $D^{\text{BD}}$ and $D^{\text{QBC}}$, BD uses $M$ times more feedback bits per user than ZFC.} the CBSUS algorithm in \cite{Shen2006a} to include the average interference due to quantization.

\begin{figure}
\begin{center}
\includegraphics[width=86mm]{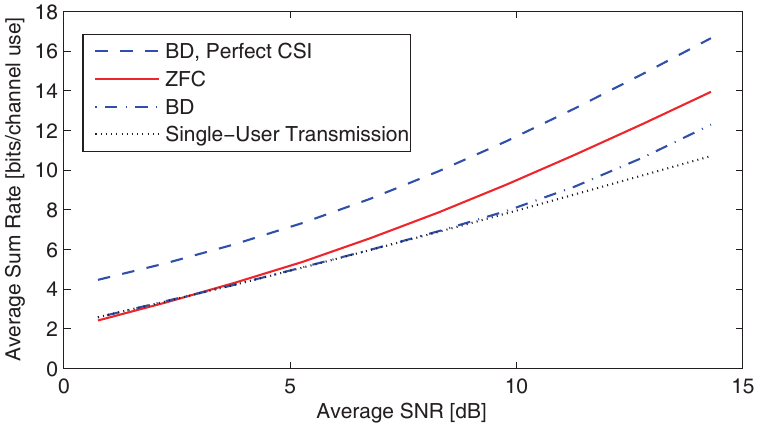} \vskip-2mm
\caption{The average achievable sum rate with BD and ZFC, quantized CSI feedback, $N=4$ transmit antennas, $M=2$ receive antennas, uncorrelated channels, and varying SNR.
The number of feedback bits is scaled with the transmit power according to Corollary \ref{cor_multiplexing_gain_quantized} and \cite[Eq.~(17)]{Ravindran2008a}.}\label{figure_RVQ_scaling}
\end{center} \vskip-5mm
\end{figure}

First, we compare BD (having either quantized or perfect CSI) with quantized ZFC using MESC-MMSE combining \cite{Trivellato2008a} and with single-user SVD-based transmission (to a randomly selected user). The quantized effective channels are obtained from 8 users under ZFC, while the entire channels are quantized for 4 users under BD. The average achievable sum rate is shown in Fig.~\ref{figure_RVQ_scaling} as a function of the average SNR. At low SNRs, quantized BD only selects one user and performs similar to single-user transmission. As two data streams are transmitted to the selected user, both strategies are slightly better than ZFC in this regime. But quantized ZFC quickly improves with SNR and becomes the method of choice at practical SNRs. The simulation was stopped at $P=14.3$ dB where BD requires feedback of 22 bits per user, meaning that the best codeword is selected in a codebook with over a million entries.\footnote{An approach to emulate RVQ for very large random codebooks was proposed in \cite{Ravindran2008a}, but this does not change the fact that the quantization complexity becomes infeasible much faster under BD than under ZFC.} BD is therefore suboptimal both in terms of sum rate and computational complexity.

\begin{figure}
\begin{center}
\includegraphics[width=86mm]{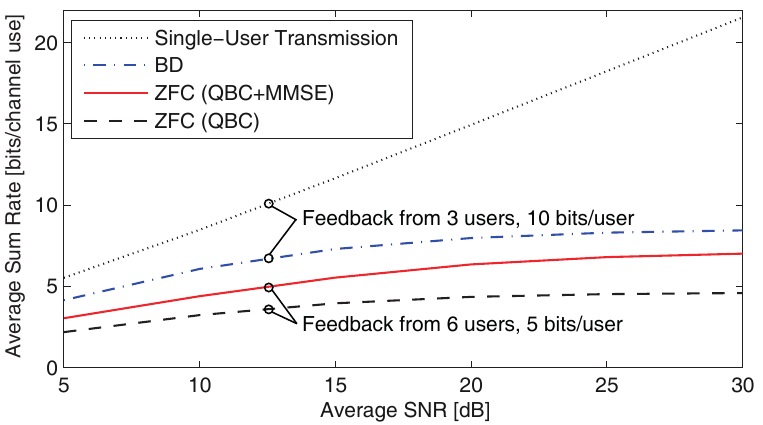} \vskip-2mm
\caption{Comparison of single-user transmission, BD, and different forms of ZFC under quantized CSI feedback. The scenario is the same as in \cite[Fig.~6]{Ravindran2008a}, where the superior single-user strategy was not included.}\label{figure_RVQ}
\end{center} \vskip-5mm
\end{figure}

This observation stands in contrast to the numerical results in \cite{Ravindran2008a}, where BD clearly beats ZFC under quantized CSI. To explain the difference, we repeat the simulation in \cite[Fig.~6]{Ravindran2008a} with $N=6$ transmit antennas and $M=2$ receive antennas.
In this simulation, the RVQ codebooks contain 10 bits/user under BD and 5 bits/user under ZFC. The achievable sum rate is shown in Fig.~\ref{figure_RVQ} for the quantized BD
approach in \cite{Ravindran2008a} and the ZFC-QBC approach in \cite{Jindal2008a}. We have also included: 1) an improved version of ZFC-QBC where the MMSE receive combiner is applied during transmission; and 2) single-user SVD-based transmission to a randomly selected user.
Our simulation confirms that BD is better than ZFC in this scenario, but the difference becomes much smaller when the MMSE combiner is applied.
However, none of these strategies should be used in this scenario since single-user transmission is vastly superior. The explanation is that the number of feedback bits is fixed at a number that only satisfies/exceeds the feedback scaling law in \eqref{eq_total_bits} and \cite[Eq.~(17)]{Ravindran2008a} at low SNRs (cf. \cite[Fig.~2]{Ravindran2008a}), while the strict interference mitigation in BD and ZFC is only practically meaningful at high SNR. The observation in \cite{Ravindran2008a} is thus misleading and does not contradict the superiority of ZFC under proper feedback loads.

Conclusions from the mathematical and numerical analysis are summarized in the next section.

\section{Conclusion}

This paper analyzed how to divide data streams among users in a downlink system with many multi-antenna users; should few users be allocated many streams, or many users be allocated few streams? New and generalized analytic results were obtained to study this tradeoff under spatial correlation, user selection, heterogeneous user channel conditions, and practical CSI acquisition.

The main conclusion is that sending one stream per selected user and exploiting receive combining is the best choice under realistic conditions. This is good news as it reduces the hardware requirements at the users, compared with multi-stream multiplexing, and enables computationally efficient resource allocation as in \cite{Bjornson2013d}. The result is explained by a stronger resilience towards spatial correlation and larger benefit from user selection. To arrive at alternative conclusions, one has to consider a scenario with heterogeneous user conditions with either perfect CSI (unrealistic) or where CSI is only acquired for the strongest users (destroys coverage and fairness). It should however be noted that if only very inaccurate CSI can be acquired, then inter-user interference will limit performance thus making single-user transmission advantageous.

\appendices

\section{Collection of Lemmas}
\label{app_collection_of_lemmas}

This appendix contains two lemmas that are essential for proving the theorems of this paper.
The first result shows how spatial correlation at the receiver affects the channel directions.
\begin{lemma} \label{lemma_row_space}
Let $\vect{A} \succ \vect{0}_M$ be any Hermitian positive-definite matrix and let $\widetilde{\vect{H}} \in \mathbb{C}^{M \times N}$ be an arbitrary matrix. Then,
$\spa(\widetilde{\vect{H}})=\spa(\vect{A} \widetilde{\vect{H}})$, where $\spa(\cdot)$ denotes the row space.
\end{lemma}
\begin{IEEEproof}
Let $\vect{A}=\vect{U}_A \vect{\Lambda}_A \vect{U}_A^H$ be an eigen decomposition of $\vect{A}$. The lemma follows by observing that $\vect{U}_A$ only rotates the basis vectors of the row space and $\vect{\Lambda}_A$ scales the rows without affecting their span.
\end{IEEEproof}

The second result generalizes the bounding of performance loss under imperfect CSI in \cite{Jindal2008a,Ravindran2008a}.

\begin{lemma} \label{lemma_inequality_rate_difference}
Let $\vect{W}_{k}, \widetilde{\vect{W}}_{k}$ be isotropically distributed on the Grassmannian manifold $\mathcal{G}_{N,d_k}$ and independent of $\vect{H}_k$, then
\begin{equation}
\begin{split}
\mathbb{E} \Big\{ &\log_2  \det \!\Big( \vect{I}_{d_k} \!+\! \frac{P}{N} \widetilde{\vect{C}}_k^H \vect{H}_k  \vect{W}_{k} \vect{W}_{k}^H \vect{H}_k^H \widetilde{\vect{C}}_k \Big) \Big\} \\
-&\mathbb{E} \left\{ \log_2 \! \frac{\det \!\Big( \vect{I}_{d_k} \!+\! \frac{P}{N}  \fracSum{\ell} \widetilde{\vect{C}}_k^H \vect{H}_k  \widetilde{\vect{W}}_{\ell} \widetilde{\vect{W}}_{\ell}^H \vect{H}_k^H \widetilde{\vect{C}}_k \Big) \!}{ \det \!\Big( \vect{I}_{d_k} \!+\! \frac{P}{N}  \fracSum{\ell \neq k}  \widetilde{\vect{C}}_k^H \vect{H}_k \widetilde{\vect{W}}_{\ell} \widetilde{\vect{W}}_{\ell}^H \vect{H}_k^H \widetilde{\vect{C}}_k \Big) \! } \right\} \\
& \leq \log_2  \det \!\Big( \vect{I}_{d_k} \!+\! \frac{P}{N}  \fracSum{\ell \neq k} \mathbb{E} \big\{ \widetilde{\vect{C}}_k^H \vect{H}_k  \widetilde{\vect{W}}_{k} \widetilde{\vect{W}}_{k}^H \vect{H}_k^H \widetilde{\vect{C}}_k \big\} \Big).
\end{split}
\end{equation}
\end{lemma}
\begin{IEEEproof}
This lemma follows from two inequalities. First,
\begin{equation} \label{eq_first_inequality}
\begin{split}
&\mathbb{E} \Big\{ \log_2  \det \!\Big( \vect{I}_{d_k} \!+\! \frac{P}{N}  \widetilde{\vect{C}}_k^H \vect{H}_k  \vect{W}_{k} \vect{W}_{k}^H \vect{H}_k^H \widetilde{\vect{C}}_k \Big) \Big\} \\
&- \mathbb{E} \Big\{ \log_2  \det \!\Big( \vect{I}_{d_k} \!+\! \frac{P}{N}  \sum_{\ell} \widetilde{\vect{C}}_k^H \vect{H}_k  \widetilde{\vect{W}}_{\ell} \widetilde{\vect{W}}_{\ell}^H \vect{H}_k^H \widetilde{\vect{C}}_k \Big) \Big\} \leq 0
\end{split}
\end{equation}
since $\widetilde{\vect{C}}_k^H \vect{H}_k  \vect{W}_{k} \vect{W}_{k}^H \vect{H}_k^H \widetilde{\vect{C}}_k$ and $\widetilde{\vect{C}}_k^H \vect{H}_k  \widetilde{\vect{W}}_k \widetilde{\vect{W}}_k^H \vect{H}_k^H \widetilde{\vect{C}}_k$ have the same distribution, and the second term contains additional positive semi-definite matrices. Second, applying Jensen's inequality on the concave function $\log_2 \det(\cdot)$ gives
\begin{equation} \label{eq_second_inequality}
\begin{split}
\mathbb{E} & \Big\{ \log_2 \det \!\Big( \vect{I}_{d_k} \!+\! \frac{P}{N} \sum_{\ell \neq k} \widetilde{\vect{C}}_k^H \vect{H}_k  \widetilde{\vect{W}}_{\ell} \widetilde{\vect{W}}_{\ell}^H \vect{H}_k^H \widetilde{\vect{C}}_k  \Big) \Big\} \\
&\leq \log_2  \det \!\Big( \vect{I}_{d_k} \!+\! \frac{P}{N} \sum_{\ell \neq k} \mathbb{E} \big\{ \widetilde{\vect{C}}_k^H \vect{H}_k  \widetilde{\vect{W}}_{\ell} \widetilde{\vect{W}}_{\ell}^H \vect{H}_k^H \widetilde{\vect{C}}_k \big\} \Big).
\end{split}
\end{equation}
The lemma follows from combining \eqref{eq_first_inequality} and \eqref{eq_second_inequality}.
\end{IEEEproof}

\section{}
\label{app_theorem_asymptotic_difference}

\textit{Proof of Theorem \ref{theorem_asymptotic_difference}:} Using \eqref{eq_sum_rate_asymptotic}, the expected asymptotic difference is
\begin{equation} \label{eq_asymptotic_difference1}
\bar{\beta}_{\text{BD-ZFC}} =  \mathbb{E}\left\{ \log_2 \left( \frac{\prod_{k \in \mathcal{S}^{\text{BD}}} \det (\vect{H}_k \vect{W}_k^{\text{BD}} \vect{W}_k^{\text{BD},H} \vect{H}_k^H)}
 { \prod_{\ell \in \mathcal{S}^{\text{ZFC}}}  |\tilde{\vect{c}}_{\ell}^H \vect{H}_{\ell} \vect{w}^{\text{ZFC}}_{\ell}|^2} \right) \right\}.
\end{equation}
The direction $\frac{\tilde{\vect{c}}_{\ell}^H \vect{H}_{\ell}}{\| \tilde{\vect{c}}_{\ell}^H \vect{H}_{\ell}\|_2}$ is isotropically distributed on the unit sphere, according to Lemma \ref{lemma_distribution_effective_channel}. This enables us to rewrite \eqref{eq_asymptotic_difference1} as
\begin{equation} \label{eq_asymptotic_difference_rewritten}
\begin{split}
\bar{\beta}_{\text{BD-ZFC}}   =& \mathbb{E}\left\{ \log_2 \left( \frac{\prod_{k \in \mathcal{S}^{\text{BD}}} \det (\widetilde{\vect{H}}_k \vect{W}_k^{\text{BD}} \vect{W}_k^{\text{BD},H} \widetilde{\vect{H}}_k^H)}
 { \prod_{k \in \mathcal{S}^{\text{ZFC}}}  |\widetilde{\vect{h}}_k^H \vect{w}^{\text{ZFC}}_{k}|^2} \right) \right\} \\ &+ \sum_{k \in \mathcal{S}^{\text{BD}}} \log_2 \det(\vect{R}_{R,k}) - \sum_{\ell \in \mathcal{S}^{\text{ZFC}}} \mathbb{E} \{ z_{\ell} \}
\end{split}
\end{equation}
where $z_{\ell} = \mathbb{E} \left\{ \log_2 \left( \frac{ \| \tilde{\vect{c}}_{\ell}^H \vect{H}_{\ell} \|_2^2 }{\| \widetilde{\vect{h}}_{\ell}\|_2^2} \right) \right\}$ and $\widetilde{\vect{h}}_{\ell} \sim \mathcal{CN}(\vect{0},\vect{I}_N)$.\footnote{The vector $\widetilde{\vect{h}}_{\ell}$ is correlated with $\tilde{\vect{c}}_{\ell}^H \vect{H}_{\ell}$ (they have the same direction), but this property does not affect the proof.} The first term in \eqref{eq_asymptotic_difference_rewritten} equals the first term in \eqref{eq_asymptotic_difference} by applying \cite[Theorem 3]{Lee2007a}. The cited theorem was stated for uncorrelated channels, but can be applied in our scenario since $\vect{W}_k^{\text{BD}}$ is not affected by the receive-side correlation matrices $\vect{R}_{R,\ell} \,\, \forall \ell \in \mathcal{S}^{\text{ZFC}}$; see Lemma \ref{lemma_row_space} in Appendix \ref{app_collection_of_lemmas}. Two bounds on $z_{\ell}$ are given in the theorem. The lower bound is achieved by the suboptimal choice of $\tilde{\vect{c}}_{\ell}$ as the dominating eigenvector of $\vect{R}_{R,k}$; this makes  $\tilde{\vect{c}}_{\ell}^H \vect{H}_{\ell} \sim \mathcal{CN}(\vect{0},\lambda_{\ell,M} \vect{I}_N)$. The upper bound is achieved from Lemma \ref{lemma_distribution_effective_channel} by applying Jensen's inequality and computing
$\mathbb{E} \{ \log_2 ( \| \widetilde{\vect{h}}_{\ell}\|_2^2 )\} = \frac{\psi(N)}{\log_e(2)}$ using standard integration.

\section{}
\label{app_theorem_impact_of_scheduling}

\textit{Proof of Theorem \ref{theorem_impact_of_scheduling}:}
We begin with BD and assume that there are $K$ candidates to become the new user $k$, $\mathcal{K}=\{1,\ldots,K\}$, while the other users in $\mathcal{S}^{\text{BD}}$ are fixed. Since $|\mathcal{S}^{\text{BD}}|=\frac{N}{M}$, all available degrees of freedom are consumed by the interference cancelation. The precoding matrix $\vect{W}_k^{\text{BD}}$ is therefore completely determined by the common null space of the co-users' channels and fixed in this proof.

Minimizing \eqref{eq_expected_precoding_loss} corresponds to finding the user $\ell \in \mathcal{K}$ with the row space of $\vect{H}_{\ell}$ most compatible with $\vect{W}_k^{\text{BD}}$. For a user candidate $\ell \in \mathcal{K}$, we can lower bound \eqref{eq_expected_precoding_loss} as
\begin{equation} \label{eq_expected_loss_BD_boundingsteps}
\begin{split}
- &\mathbb{E}\{ \log_2 \det (\vect{B}_{\ell} \vect{W}_k^{\text{BD}} \vect{W}_k^{\text{BD},H} \vect{B}_{\ell}^H) \} \\
&= - M \mathbb{E}\{ \log_2 (\det (\vect{B}_{\ell} \vect{W}_k^{\text{BD}} \vect{W}_k^{\text{BD},H} \vect{B}_{\ell}^H)^{1/M}) \} \\
&\geq -M \mathbb{E}\left\{ \log_2 \left(\frac{\tr(\vect{B}_{\ell} \vect{W}_k^{\text{BD}} \vect{W}_k^{\text{BD},H} \vect{B}_{\ell}^H )}{M}\right) \right\} \\
&\geq - M \log_2 \left(\frac{\mathbb{E}\{\tr(\vect{B}_{\ell} \vect{W}_k^{\text{BD}} \vect{W}_k^{\text{BD},H} \vect{B}_{\ell}^H )\} }{M}\right) \\
&= - M \log_2 \left(1+ \frac{\mathbb{E}\{\tr(\vect{B}_{\ell} \vect{W}_k^{\text{BD}} \vect{W}_k^{\text{BD},H} \vect{B}_{\ell}^H )- M\} }{M}\right).
\end{split}
\end{equation}
The first inequality is the classic inequality between arithmetic and geometric means, while the second inequality follows from applying Jensen's inequality on the convex function $-\log_2 \det(\cdot)$. The final expression in \eqref{eq_expected_loss_BD_boundingsteps} contains $M-\tr(\vect{B}_{\ell} \vect{W}_k^{\text{BD}} \vect{W}_k^{\text{BD},H} \vect{B}_{\ell}^H )$, which is the squared chordal distance between $\vect{B}_{\ell}$ and $\vect{W}_k^{\text{BD}}$.

Since the matrices $\vect{B}_{\ell}$, for $\ell \in \mathcal{K}$, are independent and isotropically distributed on the Grassmannian manifold $\mathcal{G}_{N,M}$ irrespective of the receive-side correlation (see Lemma \ref{lemma_row_space}), we can bring in results from \cite{Dai2008a} on quantization of Grassmannian manifolds using $K$ random codewords. From \cite[Theorem 4]{Dai2008a}, we have the following lower bound on the average squared chordal distance (for sufficiently large $K$):
\begin{equation} \label{eq_codebook_bounds}
\begin{split}
\min_{\ell \in \mathcal{K}} \, \mathbb{E} \Big\{ M-&\tr(\vect{B}_{\ell} \vect{W}_k^{\text{BD}} \vect{W}_k^{\text{BD},H} \vect{B}_{\ell}^H ) \Big\} \\ &\geq \frac{M(N-M)}{M(N-M)+1} c_{N,M,M,2}^{-\frac{1}{M(N-M)}} K^{-\frac{1}{M(N-M)}}
\end{split}
\end{equation}
where $c_{N,M,M,2}$ is a positive constant defined in \cite[Eq.~(8)]{Dai2008a}. Plugging \eqref{eq_codebook_bounds} into \eqref{eq_expected_loss_BD_boundingsteps} yields the lower bound for BD in the theorem.

A similar approach can be taken under ZFC (by setting $M=1$ in the derivation), but the $M$ receive antennas provide degrees of freedom to select the effective channel as the vector in the row space of $\vect{H}_{\ell}$ that minimizes the chordal distance to $\vect{w}_k^{\text{ZFC}}$. This is done by the QBC approach in \cite{Jindal2008a}, which was derived for uncorrelated channels but can be applied under receive correlation due to Lemma \ref{lemma_row_space}. We apply \cite[Lemma 1]{Jindal2008a}, which says that the minimal chordal distance is the minimum of $K$ independent $\beta(N-M,M)$-distributed random variables. This quantity can be lower bounded by taking the minimum of $K$ independent $\beta(N-M,1)$ variables and further lower bounded by the quantization bound in \cite[Theorem 4]{Dai2008a}:
\begin{equation} \label{eq_codebook_bounds_ZF}
\begin{split}
\min_{\ell \in \mathcal{K}} \, \mathbb{E} \left\{ \! 1 \! - \! \left|\frac{\vect{h}_{\ell}^H \vect{w}_k^{\text{ZFC}} }{\| \vect{h}_{\ell}\|_2}  \right|^2 \!\right\} \!\geq\! \frac{(N\!-\!M) K^{-\frac{1}{(N-M)}}}{(N\!-\!M)+1} c_{N-M+1,1,1,2}^{-\frac{1}{(N-M)}}
\end{split}
\end{equation}
where $c_{N-M+1,1,1,2}$ is a positive constant defined in \cite[Eq.~(8)]{Dai2008a}. Plugging \eqref{eq_codebook_bounds_ZF} into \eqref{eq_expected_loss_BD_boundingsteps} for $M=1$ yields the lower bound for ZFC in the theorem.

\section{}
\label{app_theorem_rateloss_BD}

\textit{Proof of Theorem \ref{theorem_rateloss_BD}:} Using Lemma \ref{lemma_row_space}, the row space of the correlated channel $\vect{H}_k = \vect{R}^{1/2}_{R,k} \widetilde{\vect{H}}_k$ is the same as for the uncorrelated channel $\widetilde{\vect{H}}_k$. Consequently, $\bar{\vect{W}}_{k}^{\text{BD}}$ will be isotropically distributed
on the Grassmannian manifold $\mathcal{G}_{N,M}$, just as proved for uncorrelated channels in \cite[Theorem 1]{Ravindran2008a}.
The performance loss can therefore be bounded using Lemma \ref{lemma_inequality_rate_difference} and it only remains to
characterize $\mathbb{E} \{ \vect{H}_k \bar{\vect{W}}^{\text{BD}}_{\ell} \bar{\vect{W}}_{\ell}^{\text{BD},H} \vect{H}_k^H \}$
for $\ell \neq k$. Observe that
\begin{equation}
\begin{split}
\mathbb{E}  \{ \vect{H}_k &\bar{\vect{W}}_{\ell}^{\text{BD}}  \bar{\vect{W}}_{\ell}^{\text{BD},H} \vect{H}_k^H \} \\
&=\vect{R}^{1/2}_{R,k} \mathbb{E} \{ \vect{L}_k \vect{Q}_k \bar{\vect{W}}_{\ell}^{\text{BD}} \bar{\vect{W}}_{\ell}^{\text{BD},H} \vect{Q}_k^H \vect{L}_k^H  \} \vect{R}^{1/2,H}_{R,k}
\end{split}
\end{equation}
using that $\vect{H}_k = \vect{R}^{1/2}_{R,k} \widetilde{\vect{H}}_k = \vect{R}^{1/2}_{R,k} \vect{L}_k \vect{Q}_k$, where $\vect{L}_k \in \mathbb{C}^{M \times M}$ is the lower triangular matrix and  $\vect{Q}_k \in \mathbb{C}^{M \times N}$ is the semi-unitary matrix in an LQ decomposition of $\widetilde{\vect{H}}_k$.
Observe that $\vect{L}_k$ and $\vect{Q}_k$ are independent, thus we can calculate their expectations sequentially as
\begin{equation} \label{eq_calculating_inner_expectations}
\begin{split}
\mathbb{E} \{\vect{L}_k &\vect{Q}_k \bar{\vect{W}}_{\ell}^{\text{BD}}  \bar{\vect{W}}_{\ell}^{\text{BD},H} \vect{Q}_k^H \vect{L}_k^H  \} \\
&= \frac{D^{\text{BD}}}{N-M} \mathbb{E} \{\vect{L}_k \vect{I}_M \vect{L}_k^H  \} = \frac{N D^{\text{BD}}}{N-M} \vect{I}_M.
\end{split}
\end{equation}
The first equality follows from \cite[Eq.~(43)--(45)]{Ravindran2008a}, while the second follows from
$\mathbb{E} \{ \vect{L}_k \vect{L}_k^H \} = N \vect{I}_M$ (since $\mathbb{E} \{ \widetilde{\vect{H}}_k \widetilde{\vect{H}}_k^H \} = N \vect{I}_M$).
Plugging \eqref{eq_calculating_inner_expectations} into Lemma \ref{lemma_inequality_rate_difference} yields
\begin{equation}
\Delta^{\text{BD}} \leq  \log_2 \det \left( \vect{I}_{M} +\frac{P}{N} \!\left(\frac{N}{M}-1 \right) \!\frac{N D^{\text{BD}}}{N-M} \vect{R}_{R,k} \right)
\end{equation}
from which \eqref{eq_bounding_QBD_loss_theorem} follows directly. The approximate expression for $ D^{\text{BD}}$ is given in \cite[Eq.~(26)]{Ravindran2008a}.

\section{}
\label{app_theorem_rateloss_ZF}

\textit{Proof of Theorem \ref{theorem_rateloss_ZF}:} This proof follows along the lines of \cite[Theorem 1]{Jindal2008a}, with the difference that 1) we have spatial correlation at the receiver; and 2) we use QBC also under perfect CSI. Using Lemma \ref{lemma_row_space}, we observe that the row space of the correlated channel $\vect{H}_k = \vect{R}^{1/2}_{R,k} \widetilde{\vect{H}}_k$ is the same as for the uncorrelated channel $\widetilde{\vect{H}}_k$. Since the gain of the effective channel is ignored in \eqref{eq_QBC_receiver}, the error-minimizing codeword is the same as for uncorrelated channels and we can apply \cite[Lemma 2]{Jindal2008a} to conclude that the direction of the effective channel $\vect{h}_k = \vect{H}_k^H \tilde{\vect{c}}^{\text{QBC}}_k$ is isotropically distributed. The beamforming vector $\bar{\vect{w}}_{k}^{\text{ZFC}}$ is independent of $\vect{h}_k$ and also isotropic, thus the performance loss can be bounded using Lemma \ref{lemma_inequality_rate_difference}. It only remains to characterize $\mathbb{E} \{ | \vect{h}_k^H \bar{\vect{w}}^{\text{ZFC}}_{\ell} |^2 \} = \mathbb{E} \{ \| \vect{h}_k \|_2^2 \} \mathbb{E} \{ | \frac{ \vect{h}_k^H}{\|\vect{h}_k\|_2} \bar{\vect{w}}^{\text{ZFC}}_{\ell} |^2 \}$ for $\ell \neq k$. The second factor equals $\frac{D^{\text{QBC}}}{N-1}$ using \cite[Lemma 2]{Jindal2006a} and \cite[Eq.~(17)]{Jindal2008a}, while computing the average norm $\mathbb{E} \{ \| \vect{h}_k \|_2^2 \}$ of the effective channel is nontrivial.  To enable reuse of results from \cite{Jindal2008a}, let $\tilde{\vect{c}}^{\text{U-QBC}}_k$ be the QBC for the uncorrelated channel $\widetilde{\vect{H}}_k$ and observe that
\begin{equation}
\tilde{\vect{c}}^{\text{QBC}}_k = \frac{\vect{R}^{-1/2}_{R,k} \tilde{\vect{c}}^{\text{U-QBC}}_k}{\| \vect{R}^{-1/2}_{R,k} \tilde{\vect{c}}^{\text{U-QBC}}_k \|_2}.
\end{equation}
We can therefore express the effective channel as
\begin{equation}
\begin{split}
\vect{h}_k =\vect{H}_k^H \tilde{\vect{c}}^{\text{QBC}}_k =\vect{H}_k^H \frac{\vect{R}^{-1/2}_{R,k} \tilde{\vect{c}}^{\text{U-QBC}}_k}{\| \vect{R}^{-1/2}_{R,k} \tilde{\vect{c}}^{\text{U-QBC}}_k \|_2} = \frac{\widetilde{\vect{H}}_k^H \tilde{\vect{c}}^{\text{U-QBC}}_k}{\| \vect{R}^{-1/2}_{R,k} \tilde{\vect{c}}^{\text{U-QBC}}_k \|_2}
\end{split}
\end{equation}
and its squared norm will be
\begin{equation} \label{eq_norm_efficient_channel}
\|\vect{h}_k\|_2^2 = \|\widetilde{\vect{H}}_k^H \tilde{\vect{c}}^{\text{U-QBC}}_k\|_2^2  \frac{1}{\tilde{\vect{c}}^{\text{U-QBC},H}_k \vect{R}^{-1}_{R,k} \tilde{\vect{c}}^{\text{U-QBC}}_k}.
\end{equation}
The first factor is the same as under uncorrelated fading and satisfies
$\mathbb{E} \{ \|\widetilde{\vect{H}}_k^H \tilde{\vect{c}}^{\text{U-QBC}}_k\|_2^2 \}= N-M+1$ (see \cite[Lemma 4]{Jindal2008a}), while the second factor depends on $\vect{R}_{R,k}$.
Since both the quantization codebook and $\widetilde{\vect{H}}_k$ are isotropically distributed, $\tilde{\vect{c}}^{\text{U-QBC}}_k$ is also isotropic and the two terms in
\eqref{eq_norm_efficient_channel} are independent. To characterize the second term, observe that $\tilde{\vect{c}}^{\text{U-QBC}}_k$ can be viewed as a normalized uncorrelated circular-symmetric complex Gaussian vector. By using that the eigenvectors of $\vect{R}_{R,k}$ are not affecting the distribution and that squared magnitudes of $\mathcal{CN}(0,1)$-variables are exponentially distributed \cite{Hammarwall2008a}, we conclude that the second term of \eqref{eq_norm_efficient_channel} has the same distribution as
\begin{equation}
\frac{\sum_{i=1}^{M} \xi_i}{\sum_{i=1}^{M} \frac{\xi_i}{\lambda_{k,i}}}
\end{equation}
for some independent exponentially distributed $\xi_i \sim \mathrm{Exp}(1)$. For any $a$ such that $\lambda_{k,m} \leq a \leq \lambda_{k,m+1}$, we can write the CDF as
\begin{equation}
\begin{split}
\mathrm{Pr}&\left\{ \frac{\sum_{i=1}^{M} \xi_i}{\sum_{i=1}^{M} \frac{\xi_i}{\lambda_{k,i}}} \leq a \right\} \\ &= \mathrm{Pr}\Bigg\{
\sum_{i=1}^{m} \underbrace{\left( \frac{a}{\lambda_{k,i}}-1\right)}_{\geq 0} \xi_i - \! \sum_{i=m+1}^{M} \underbrace{\left( 1-\frac{a}{\lambda_{k,i}}\right)}_{\geq 0} \xi_i
\geq 0 \Bigg\}.
\end{split}
\end{equation}
This is the difference of two sums of exponentially distributed variables (with distinct positive variances). The PDF of each sum is characterized by \cite[Theorem 4]{Hammarwall2008a} and by calculating their convolution and integrating over all positive values, we achieve the CDF
\begin{equation}
\begin{split}
\mathrm{Pr}&\left\{ \frac{\sum_{i=1}^{M} \xi_i}{\sum_{i=1}^{M} \frac{\xi_i}{\lambda_{k,i}}} \leq a \right\} \\& =
\sum_{n=1}^m \sum_{t=m+1}^{M} \frac{ (\mu_n-a^{-1})^{m} (a^{-1}-\mu_t)^{M-m-1} }{(\mu_n-\mu_t) \condProd{i=1}{i \neq n}{m} (\mu_n-\mu_i) \condProd{j=m+1}{j \neq l}{M} \!\!(\mu_j-\mu_t)  }
\end{split}
\end{equation}
using the simplifying notation $\mu_{n} = \frac{1}{\lambda_{k,n}}$. The corresponding mean value is achieved from the CDF by simply taking the derivative and sum up the mean values over each $a$-interval. By multiplying the mean value expression with $N-M+1$ (i.e., the contribution of the first part in \eqref{eq_norm_efficient_channel}), we achieve the expression for $G_k$.

\section{}
\label{app_theorem_rateloss_BD_EST}

\textit{Proof of Theorem \ref{theorem_rateloss_BD_EST}:} The proof follows along the lines of Theorem \ref{theorem_rateloss_BD}, but we consider CSI estimation errors instead of quantization errors. First, observe that both $\vect{W}_{\ell}^{\text{BD}}$ and $\widehat{\vect{W}}_{\ell}^{\text{BD}}$ are isotropically distributed
on the Grassmannian manifold $\mathcal{G}_{N,M}$ (since receive-side correlation is not affecting the row space of $\vect{H}_k$ and $\widehat{\vect{H}}_k$; see Lemma \ref{lemma_row_space}).
The performance loss can therefore be bounded using Lemma \ref{lemma_inequality_rate_difference} and it only remains to
characterize $\mathbb{E} \{ \vect{H}_k \widehat{\vect{W}}^{\text{BD}}_{\ell} \widehat{\vect{W}}_{\ell}^{\text{BD},H} \vect{H}_k^H \}$
for $\ell \neq k$. From \eqref{eq_MMSE_estimator} we have
\begin{equation} \label{eq_true_estimated_channel}
\vect{H}_k = \widehat{\vect{H}}_k + \vect{R}_{E,k}^{1/2}\widetilde{\vect{E}}_k
\end{equation}
where the second term is the estimation error, $\vect{R}_{E,k} = \left(\vect{R}_{R,k}^{-T} + \frac{\vect{T}_k^H \vect{T}_k}{\sigma^2} \right)^{-1}$, and $\widetilde{\vect{E}}_k$ has $\mathcal{CN}(0,1)$-entries.
By using that $\widehat{\vect{H}}_k \widehat{\vect{W}}_{\ell}^{\text{BD}} = \vect{0}$ for $\ell \neq k$, we achieve
\begin{equation}
\begin{split}
 \mathbb{E}  \{ \vect{H}_k \widehat{\vect{W}}_{\ell}^{\text{BD}}  \widehat{\vect{W}}_{\ell}^{\text{BD},H} \vect{H}_k^H \}
= \vect{R}_{E,k}^{1/2}
\mathbb{E}  \{ \widetilde{\vect{E}}_k \widehat{\vect{W}}_{\ell}^{\text{BD}}  \widehat{\vect{W}}_{\ell}^{\text{BD},H} \widetilde{\vect{E}}_k^H \}
\vect{R}_{E,k}^{1/2}
\end{split}
\end{equation}
where $\mathbb{E} \{ \widetilde{\vect{E}}_k \widehat{\vect{W}}_{\ell}^{\text{BD}}  \widehat{\vect{W}}_{\ell}^{\text{BD},H} \widetilde{\vect{E}}_k^H \}= M\vect{I}_M$ since $\widetilde{\vect{E}}_k$ is complex Gaussian and independent of $\widehat{\vect{W}}_{\ell}^{\text{BD}}$.
Therefore, $\mathbb{E} \{ \vect{H}_k \widehat{\vect{W}}^{\text{BD}}_{\ell} \widehat{\vect{W}}_{\ell}^{\text{BD},H} \vect{H}_k^H \} = M \left( \vect{R}_{R,k}^{-T} + \frac{\vect{T}_k^H \vect{T}_k}{\sigma^2} \right)^{-1}$.

\section{}
\label{app_lemma_distribution_effective_channel}

\textit{Proof of Lemma \ref{lemma_distribution_effective_channel}:} Observe that $\vect{H} = \vect{R}^{1/2}_{R} \widetilde{\vect{H}}$ has the same distribution as $\vect{H} \vect{U}$ for any unitary matrix $\vect{U}$.
Thus, we can rotate $\vect{h}$ arbitrarily without changing the statistics, meaning that $\frac{\vect{h}}{\|\vect{h}\|_2}$ must be isotropically distributed.
Next, note that $\| \vect{h}\|_2^2 = \|\tilde{\vect{c}}^H \vect{H} \vect{U}\|_2^2 = \|\tilde{\vect{c}}^H \vect{H}\|_2^2$, thus unitary rotations will not affect the effective channel gain meaning that the direction and the channel gain are statistically independent.
$\| \vect{h}\|_2^2$ is the dominating eigenvalue of the correlated complex Wishart matrix $\vect{H} \vect{H}^H \in \mathcal{W}_M(N,\vect{R}_{R})$. The expectation in \eqref{eq_mean_value_effective_channel} achieved directly from \cite[Theorem 3]{Bjornson2008d} or by using the moment generating function in \cite{Mckay2006b} (which gives an equivalent expression that looks slightly different).

\section{}
\label{app_theorem_rateloss_ZF_EST}

\textit{Proof of Theorem \ref{theorem_rateloss_ZF_EST}:} This theorem is proved in the same way as Theorem \ref{theorem_rateloss_BD_EST}. The only notable difference is that we use the effective channel $\vect{h}_k$, which has a single effective receive antenna, instead of the original channel $\vect{H}_k$. The effective channel is zero-mean and has an average channel gain $\mathbb{E}\{\|\vect{h}_k\|_2^2\}$ given by \eqref{eq_mean_value_effective_channel}. Thus, the effective channel and its channel estimate is related as $\vect{h}_k^H= \widehat{\vect{h}}_k^H + \left( \frac{1}{\mathbb{E}\{\|\vect{h}_k\|_2^2\}} + \frac{\Psi}{\sigma^2} \right)^{-1/2} \widetilde{\vect{e}}_k^H$, where $\widetilde{\vect{e}}_k \sim \mathcal{CN}(\vect{0},\vect{I}_N)$.

\section{}
\label{app_cor_multiplexing_gain_estimated}

\textit{Proof of Corollary \ref{cor_multiplexing_gain_estimated}:} The sufficiency is easily achieved from Theorem \ref{theorem_rateloss_BD_EST} and Theorem \ref{theorem_rateloss_ZF_EST}. To obtain the necessity with BD, consider the interference term
$\vect{H}_k \widehat{\vect{W}}_{\ell}^{\text{BD}} \widehat{\vect{\Upsilon}}_{\ell} \widehat{\vect{W}}_{\ell}^{\text{BD},H} \vect{H}_k^H = \vect{R}_{E,k}^{1/2}
\widetilde{\vect{E}}_k \widehat{\vect{W}}_{\ell}^{\text{BD}}  \widehat{\vect{W}}_{\ell}^{\text{BD},H} \widetilde{\vect{E}}_k^H \vect{R}_{E,k}^{1/2}$, where the true and estimated are related as in \eqref{eq_true_estimated_channel}. This term must be bounded as $P \rightarrow \infty$. Since the product $\widetilde{\vect{E}}_k \widehat{\vect{W}}_{\ell}^{\text{BD}}$ is non-zero almost surely and $ \widehat{\vect{\Upsilon}}_{\ell} \rightarrow \frac{P}{M |\mathcal{S}^{\text{BD}}|} \vect{I}_M$, it is necessary that $\frac{1}{P} \vect{R}_{E,k}$ has bounded elements. This makes \eqref{eq_training_law} a necessary condition. The proof for ZFC is analogous.

\bibliographystyle{IEEEtran}
\bibliography{IEEEabrv,refs}

\begin{IEEEbiography}[{\includegraphics[width=1in,height=1.25in,clip,keepaspectratio]{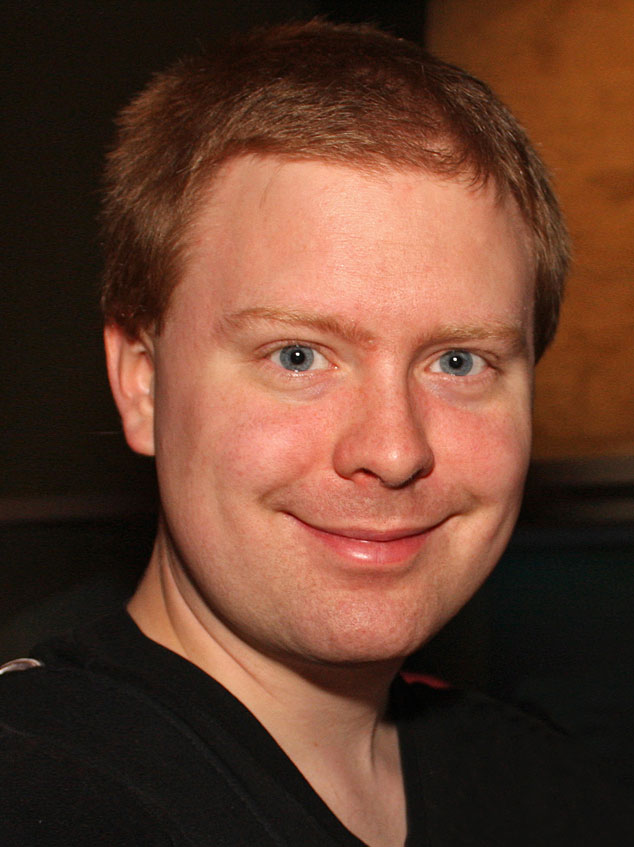}}]{Emil Bj\"ornson}
(S'07-M'12) was born in Malm\"o, Sweden, in 1983. He received the M.S.~degree in Engineering Mathematics from Lund University, Lund, Sweden, in 2007. He received the Ph.D. degree in Telecommunications from the Signal Processing Lab at KTH Royal Institute of Technology, Stockholm, Sweden, in 2011. He is the first author of the book ``Optimal Resource Allocation in Coordinated Multi-Cell Systems'' published in 2013.

Dr.~Bj\"ornson was one of the first recipients of the International Postdoc Grant from the Swedish Research Council. This grant is currently funding a joint postdoctoral research fellowship at the Alcatel-Lucent Chair on Flexible Radio, SUPELEC, Paris, France, and the Signal Processing Lab at KTH Royal Institute of Technology, Stockholm, Sweden. His research interests include multi-antenna communications, resource allocation, random matrix theory, estimation theory, stochastic signal processing, and optimization.

For his work on optimization of multi-cell MIMO communications, he received a Best Paper Award at the 2009 International Conference on Wireless Communications \& Signal Processing (WCSP'09) and a Best Student Paper Award at the 2011 IEEE International Workshop on Computational Advances in Multi-Sensor Adaptive Processing (CAMSAP'11).
\end{IEEEbiography}

\begin{IEEEbiography}[{\includegraphics[width=1in,height=1.25in,clip,keepaspectratio]{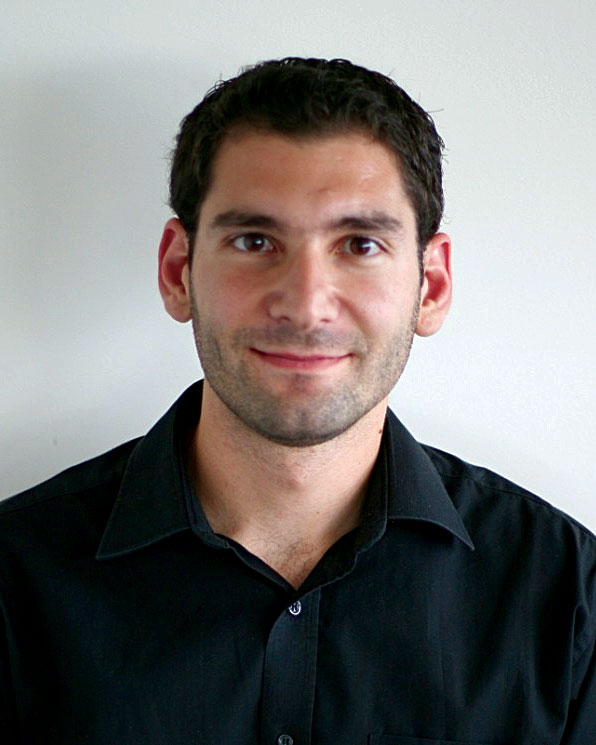}}]{Marios Kountouris}
(S'04-M'08) received the
Diploma in Electrical and Computer Engineering
from the National Technical University of Athens,
Greece, in 2002 and the M.Sc.~and Ph.D.~degrees
in Electrical Engineering from the Ecole Nationale
Sup\'erieure des T\'el\'ecommunications (Telecom Paris-
Tech), France, in 2004 and 2008, respectively. His
doctoral research was carried out at Eurecom Institute,
France, funded by France Telecom - Orange
Labs, France. From February 2008 to May 2009,
he has been with the Department of Electrical and
Computer Engineering at the University of Texas at Austin, USA, as a
postdoctoral research associate, working on wireless ad hoc networks under
DARPA's ITMANET program. Since June 2009 he has been with the Department
of Telecommunications at SUPELEC (Ecole Sup\'erieure D'Electricit\'e),
France where he is currently an Assistant Professor. His research interests
include multiuser multi-antenna communications, heterogeneous wireless networks,
interference modeling, and network information theory.
He is currently an Editor for the EURASIP Journal on Wireless Communications and Networking and Vice Chair of IEEE SIG on Green Cellular Networks.
He received the Best Paper Award in Communication Theory Symposium at the IEEE
Globecom conference in 2009 and the 2012 IEEE SPS Signal Processing
Magazine Award. He is a Member of the IEEE and a Professional Engineer
of the Technical Chamber of Greece.
\end{IEEEbiography}

 \vskip-10mm

\begin{IEEEbiography}[{\includegraphics[width=1in,height=1.25in,clip,keepaspectratio]{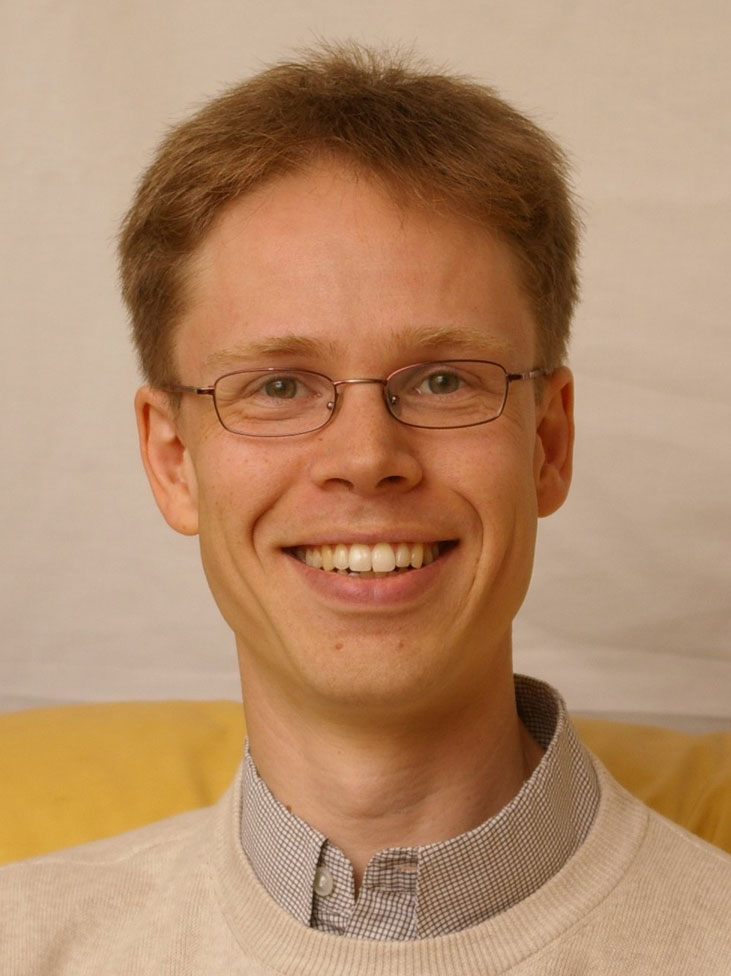}}]{Mats Bengtsson}
(M'00-SM'06) received the M.S. degree in computer science from Link\"oping University, Link\"oping, Sweden, in 1991 and the Tech.~Lic.
and Ph.D. degrees in electrical engineering from the KTH Royal Institute of
Technology, Stockholm, Sweden, in 1997 and 2000, respectively.

From 1991 to 1995, he was with Ericsson Telecom AB Karlstad. He currently holds a position as Associate Professor at the Signal Processing Laboratory, School of Electrical Engineering, KTH. His research interests include statistical signal processing and its applications to communications, multi-antenna processing, cooperative communication, radio resource management, and propagation channel modelling. Dr.~Bengtsson served as Associate Editor for the IEEE Transactions on Signal Processing 2007-2009 and was a member of the IEEE SPCOM Technical Committee 2007-2012.
\end{IEEEbiography}

\vskip-10mm

\begin{IEEEbiography}[{\includegraphics[width=1in,height=1.25in,clip,keepaspectratio]{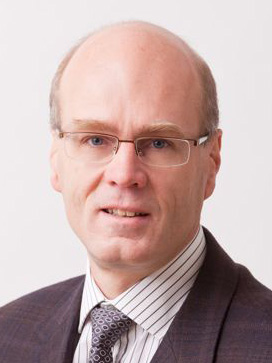}}]{Bj\"orn Ottersten}
(S'87-M'89-SM'99-F'04) was born in Stockholm, Sweden, 1961. He received the M.S. degree in electrical engineering and applied physics from Link\"oping University, Link\"oping, Sweden, in 1986. In 1989 he received the Ph.D.~degree in electrical engineering from Stanford University, Stanford, CA.
Dr.~Ottersten has held research positions at the Department of Electrical Engineering, Link\"oping University, the Information Systems Laboratory, Stanford University, the Katholieke Universiteit Leuven, Leuven, and the University of Luxembourg. During 96/97 Dr.~Ottersten was Director of Research at ArrayComm Inc, a start-up in San Jose, California based on Ottersten's patented technology. He has co-authored journal papers that received the IEEE Signal Processing Society Best Paper Award in 1993, 2001, and 2006 and 3 IEEE conference papers receiving Best Paper Awards. In 1991 he was appointed Professor of Signal Processing at the KTH Royal Institute of Technology, Stockholm. From 1992 to 2004 he was head of the department for Signals, Sensors, and Systems at KTH and from 2004 to 2008 he was dean of the School of Electrical Engineering at KTH.

Currently, Dr.~Ottersten is Director for the Interdisciplinary Centre for Security, Reliability and Trust at the University of Luxembourg. As Digital Champion of Luxembourg, he acts as an adviser to European Commissioner Neelie Kroes. Dr.~Ottersten has served as Associate Editor for the IEEE Transactions on Signal Processing and on the editorial board of IEEE Signal Processing Magazine. He is currently editor in chief of EURASIP Signal Processing Journal and a member of the editorial board of EURASIP Journal of Applied Signal Processing. Dr.~Ottersten currently serves on the IEEE Signal Processing Society Board of Governors and is a Fellow of the IEEE and EURASIP. In 2011 he received the IEEE Signal Processing Society Technical Achievement Award. He is a first recipient of the European Research Council advanced research grant. His research interests include security and trust, reliable wireless communications, and statistical signal processing.
\end{IEEEbiography}

\end{document}